\documentclass[letterpaper, 10 pt, conference]{ieeeconf}  

\IEEEoverridecommandlockouts                              

\usepackage{cite}
\usepackage{url}

\usepackage{amssymb,amsmath,amstext,amsfonts}
\usepackage{mathtools}
\usepackage{bm}  
\usepackage{siunitx}  

\usepackage{url}
\usepackage{filecontents}
\usepackage{stmaryrd}

\usepackage{graphicx}
\usepackage{color}

\usepackage{epsfig} 
\usepackage{subfig}

\graphicspath{%
  {Figures/}
  {Figures/TiKz/}
  {Figures/Ipe/}
  {Figures/XFig/}
  {Figures/Inkscape/}
  {Figures/Matlab/}
  {Figures/Scanned/}
  {Figures/Saved/}
}

\usepackage{tikz}
\usetikzlibrary{decorations.markings}
 \tikzset{every picture/.style={line width=0.75pt}} 

\usepackage[figuresright]{rotating}
\usepackage[colorinlistoftodos]{todonotes}
\usepackage[english,algo2e,algoruled,vlined,linesnumbered]{algorithm2e}   
\usepackage{enumerate}
 
\usepackage{lipsum}								
\usepackage{fancyhdr}

\usepackage[firstpageonly=true]{draftwatermark}				

\usepackage{hyperref}
\hypersetup{
    colorlinks=true,
    linkcolor=blue,
    filecolor=magenta,      
    urlcolor=cyan,
    bookmarks=true,
}
\title{\LARGE \bf Guidance Mechanism for Flexible Wing Aircraft Using Measurement-Interfaced Machine Learning Platform}

\author{Mohammed Abouheaf, Nathaniel Mailhot,  Wail Gueaieb, and Davide Spinello  
\thanks{This work was partially supported by Ontario Centers of Excellence~(OCE). Grant  number 27404.}
\thanks{Mohammed Abouheaf and Wail Gueaieb are with the School of Electrical Engineering \& Computer Science, University of Ottawa, Ottawa, Canada. Nathaniel Mailhot and Davide Spinello are with the Department of Mechanical Engineering, University of Ottawa, Ottawa, Canada. E-mail:~\{mabouhea,nmailhot,wgueaieb,dspinell\}@uottawa.ca}}

\begin{document}
\maketitle

\pagenumbering{gobble}

\DraftwatermarkOptions{%
angle=0,
hpos=0.5\paperwidth,
vpos=0.97\paperheight,
fontsize=0.012\paperwidth,
color={[gray]{0.2}},
text={
  \parbox{0.99\textwidth}{This is the accepted version of the paper. Published paper DOI: \href{http://dx.doi.org/10.1109/TIM.2020.2985553}{10.1109/TIM.2020.2985553}.\\
    \copyright\ 2020 IEEE.  Personal use of this material is permitted.  Permission from IEEE must be obtained for all other uses, in any current or future media, including reprinting/republishing this material for advertising or promotional purposes, creating new collective works, for resale or redistribution to servers or lists, or reuse of any copyrighted component of this work in other works.}},
}

\begin{abstract}
The autonomous operation of flexible-wing aircraft is technically challenging and has never been presented within literature. The lack of an exact modeling framework is due to the complex nonlinear aerodynamic relationships governed by the deformations in the flexible-wing shape, which in turn complicates the controls and instrumentation setup of the navigation system. This urged for innovative approaches to interface affordable instrumentation platforms to autonomously control this type of aircraft. This work leverages ideas from instrumentation and measurements, machine learning, and optimization fields in order to develop an autonomous navigation system for a flexible-wing aircraft. A novel machine learning process based on a guiding search mechanism is developed to interface real-time measurements of wing-orientation dynamics into control decisions. This process is realized using an online value iteration algorithm that decides on two improved and interacting model-free control strategies in real-time. The first strategy is concerned with achieving the tracking objectives while the second supports the stability of the system. A neural network platform that employs adaptive critics is utilized to approximate the control strategies while approximating the assessments of their values. An experimental actuation system is utilized to test the validity of the proposed platform. The experimental results are shown to be aligned with the stability features of the proposed model-free adaptive learning approach. 
\end{abstract}

\section{Introduction}
\label{sec:introdcution}

Flexible-wing aircraft have been capturing increasing interests due to their relatively simple mechanical structures, flexible operation, low fabrication costs, and high payload-to-mass ratio~\cite{Ochi_2017,Cook_Kilkenny_1986,Abouheaf2018Rob}.  On one side, there are no experimentally-validated dynamical models for such systems since they are characterized by highly nonlinear dynamics. Therefore, designing algorithms for the autonomous flight control of this type of aircraft is a complex process. To this end, relying on conventional methods which are based on some sort of classical mathematical models, however approximate they may be, is not an option. On another side, analytical solutions for the optimal tracking control problems often necessitate solving offline coupled differential equations simultaneously, where a subset of these equations is solved backward in-time~\cite{Lewis_2012}. Additionally, the complexity of the tracking control laws, resulting from adaptive systems, are not realizable or hard-to-implement  using digital processing units.

In this work, a coupled instrumentation-machine learning framework is proposed to solve the guidance control problem of a class of nonlinear dynamical systems with unknown dynamics, typical for flexible-wing aircraft. This process tackles difficulties associated with developing control solutions for partially- or fully model-based approaches in uncertain dynamical environments.  Further, it solves the tracking control problem in real-time with no offline or backward solution for the underlying differential equations. Simply, it provides a simplified method to compute the tracking control laws which may be both complex and hard-to-realize in digital simulation environments. The proposed scheme was tested in real-time aboard of a Raspberry~Pi equipped with an Inertial Measurement Unit (IMU) consisting of accelerometers, gyroscopes and magnetometers, in order to provide accurate orientation and motion sensing measurements (i.e., for the wing of the aircraft). A reinforcement learning process based on adaptive guided search mechanism is employed to steer the motion dynamics of a mock-up emulating the aircraft's wing movement relative to the fuselage. It employs an online value iteration process that produces real-time model-free tracking control strategies.

Flexible-wing aircraft are composed of two interacting  structures, namely wing and fuselage, which are pinned at a point where they can rotate on pitch-roll axes relative to one another~\cite{Kilkenny_1984,Kilkenny_1986,Blake_1991,cook_1994}. 
The control of this type of aircraft is challenging due to the flexibility of the wing which induces continuous aerodynamic variations and makes it difficult to model the vehicle's dynamics~\cite{Ochi_2015,Ochi_2017}. Partial theoretical models have been presented for this type of aircraft in~\cite{Sweeting1981,cook_1994,Cook1997}. The approximate aerodynamic models varied in complexity and approach. Some researchers opted to decouple the dynamics into  longitudinal and lateral motion frames to simplify the problem~\cite{Kroo_1983}. A fully decoupled aerodynamic model is developed in~\cite{Spottiswoode_2001}. In another context, a fixed-wing approach that considers aerodynamic derivatives is employed in~\cite{DE_MATTEIS_1990,De_Matteis_1991}. Additionally, a system of nonlinear dynamical equations based on a nine-degree-of-freedom dynamic model is derived in~\cite{Ochi_2015,Ochi_2017}. The navigation process of this aircraft is based on a weight-shift mechanism where the relative variations in the Center-of-Gravity (CoG) locations of the wing-fuselage systems, under kinematic and dynamic constraints at the interconnection points, steer the aircraft toward the desired orientations~\cite{cook_spottiswoode_2005,Cook_2013,AbouheafRob18}. It is shown that when a weight-shift mechanism is applied, the longitudinal stability is magnified~\cite{cook_spottiswoode_2005,De_Matteis_1991}. Also, the stability margins corresponding to the lateral motion are shown to be larger than those of fixed-wing aircraft~\cite{cook_1994}.

The tracking, navigation, and motion-guided systems employ instrumentation configurations along with theoretical advances for the underlying physical mechanisms.
A method based on radio frequency identification technology is proposed for navigating mobile robots in~\cite{Gueaieb2008}.
An optoacoustic positioning scheme that utilizes inertial and distance measurements is developed for automated monitoring of complex manual assembly operations in~\cite{Esslinger2019}. It employs a particle filter that fuses inertial navigation measurements with unilateral distances measurements to space-fixed receiving devices. 
An artificial neural network approach is proposed to fuse differential and uncompensated measurements from  global position and inertial-navigation devices  respectively in~\cite{El-Sheimy2006}. This approach avoided using Kalman filtering in order to integrate the inertial navigation and global position systems. Adaptive robust tracking control approach that uses a continuous model based estimator is employed to control a flexible air-breathing hypersonic vehicle in~\cite{Tao2019}. It employs a type-2 Fuzzy structure to approximate the unknown model dynamics and the stability features are validated using a Lyapunov theorem. A comprehensive test-bed that utilizes multi-camera for operations which involve control  and 3D tracking tasks of unmanned aerial vehicles is developed in~\cite{Deng2019}. The embedded navigation approach relies on a Proportional-Derivative (PD) control structure that receives navigation information form multi-camera system. The test-bed allowed real-time computations at \SI{100}{\Hz} using cameras with field programmable gate array (FPGA) modules where the embedded software is able to perform motion control and image processing. A vision-based tracking approach that uses particle swarm optimization and fuzzy logic scheme is developed to navigate an autonomous mobile robot in~\cite{Sharma2012}. The fuzzy tracking system is designed using a Lyapunov framework and it benefits global search capabilities of the particle swarm optimization technique. An approach based on interval analysis is developed to solve the localization problem of a mobile robot using ultrasonic sensors in~\cite{Ashokara2009}. This approach  manages the issues arising from data-associate step noticed within the classical localization problems using Kalman filtering.

Dynamic Programming (DP) solution  techniques are employed to solve different control problems~\cite{Sutton_1998,Werbos1974,Lewis_2012}. However, these frameworks degrade due to the curse of dimensionality associated with the state-action domains~\cite{Sutton_1998,Werbos1974}. Approximate Dynamic Programming (ADP) relaxed the manner at which dynamic programming problems are solved using heuristic platforms~\cite{Sutton_1998,Werbos_1990}. 
These solution forms are meant to provide computational platforms to solve the control problems using temporal difference structures~\cite{AbouheaIJCNN2013,Abouheaf2014,AbouheafCDC13}. The control problems are solved by optimizing the performance of the underlying dynamical systems using objective cost functions. Hence, the solutions for the underlying Bellman optimality or Hamilton-Jacobi-Bellman equations lead to solutions for the optimal control problems~\cite{Bellman1957,Bryson1996,Lewis_2012}. These optimal forms vary in structure, as they depend either on Bellman or Hamiltonian structures, allowing different temporal difference solution forms~\cite{AbouheafCTT2015,AbouheafIRIS17}. These solutions become more complicated in case of coupled hierarchical control systems or multi-agent structures~\cite{Basar1999,Abouheaf2014}. This is due to the existence of coupled temporal difference formulations. The ADP problems are solved using a dynamic learning environment known as Reinforcement Learning (RL)~\cite{Sutton_1998,Widrow1973,Werbos1989,Werbos_1992}. In this process, the strategies taken by the agent are either rewarded or penalized based on their value-assessments using a utility or an objective cost function. The policy-value assessment mechanism is repeated until the best strategy is found. 
%

Reinforcement learning solutions are implemented using two-step mechanisms known as value iteration and policy iteration~\cite{Sutton_1998,Widrow1973}. The first step solves the temporal difference equation, while the second approximates the optimal control policy~\cite{AbouheafCTT2015,AbouheafIRIS17}. In value iteration, the solving value function is evaluated and then the best strategy is extracted. While in policy iteration, a strategy is evaluated in the first step then a policy improvement step is introduced~\cite{Sutton_1998}. The adaptive learning solutions are developed for multi-agent systems where the communications between the nodes are accomplished using graph structures in~\cite{Abouheapolicy2017,AbouheafECC14,AbouheafIRIS17,Abouheaf2014}. These solution forms were able to solve coupled temporal difference equations in real-time using only neighborhood information. Adaptive critics are used as neural network approximation tools to implement these solutions for single- and multi-agent control problems~\cite{Bertsekas1996,Busoniu2008,Widrow1973,Werbos1989}. Each adaptive critics scheme involves two approximating neural networks; the actor network approximates the optimal control strategy while the critic network approximates the solving value function. 
%
%
A cell-mapping approach based on a reinforcement learning technique is developed for robot motion planning in~\cite{Plaza2009}. The learning mechanism does not employ a dynamical model for the robot as it builds experience-knowledge about the dynamical environment and robots dynamics. It employs a transformation based on cell-to-cell transitions in order to reduce time used to build experience about the dynamical environment.

The proposed work advances initial research investigations aimed to design intelligent flight controllers for flexible-wing aircraft using a linear actuation mechanism and machine learning process~\cite{AbouRose19}. However, this development  relied on a geometric model of the aircraft and the control strategies are mapped to actuation lengths. Herein, an experimental mock-up system that is composed of servo actuation winch motors acting on the mast and wing of a flexible-wing aircraft are employed to achieve autonomous maneuvers. In~\cite{AbouRose19}, the desired strategies are only evaluated in terms of the tracking error signals, while the proposed mechanism finds control decisions based on tracking error signals, their dynamics, and the dynamical measures of the key orientation parameters to support the overall stability of the aircraft. Finally in contrast to~\cite{AbouRose19}, this work provides a flexible and innovative model-free guided search learning process based on  a real-time value iteration scheme.  The machine learning search method opens the door to reflect the designer considerations, regarding data driven structures, in the architecture of the learning process.

The contributions of this work are four-fold. 
First, an innovative experimental platform is developed for autonomous control of flexible-wing aircraft. This is known to be a complicated task which was not realized using online model-free control systems before introducing this work.
Second, a new machine learning process is proposed. It uses real-time measurements and it conditions the optimization objectives in order to guide the search for the best control strategies in an online fashion with high success rate. 
Third, it provides ideas related to the online solutions of the optimal tracking control problems, which are solved basically in an offline fashion.   
This approach does not employ any model-dependent control strategies and it is easily integrated into an off-the-shelf computing unit, such as a Raspberry~Pi.
Finally, the research presented herein provides a flexible framework that can be generalized for controlling complex nonlinear dynamical systems of the same class as flexible-wing aircraft.

The remaining sections of the paper are arranged as follows. The operation of flexible-wing aircraft is briefly highlighted in Section~\ref{sec:operation}. The measurement scheme and the adopted real-time sensory devices are detailed in Section~\ref{sec:Inst}. The different control and optimization objectives are presented in Section~\ref{sec:control}. This is followed by the development of the machine learning process and its adaptive critics implementation in Section~\ref{sec:Machine}. The digital experimental results are analyzed in Section~\ref{sec:simulation} in order to evaluate the validity of the proposed control setup. Finally, conclusions are drawn in Section~\ref{sec:conc}.

\section{Operation of Flexible-wing Aircraft}
\label{sec:operation}

In this section, the basic kinematic relationships concerned with pitch motion control of the flexible-wing aircraft are detailed out. The two-mass system (i.e., wing and fuselage) attaches the two bodies by a hang block, a joint that connects the mast of the fuselage and the keel tube of the wing at a common point known as the hang point. The hang block is a mechanical joint which allows for two degrees of motion relative to each body. The wing system may roll freely about the longitudinal axis of the keel tube, while the keel bar may pitch about the axis which is perpendicular to both the mast and keel tubes. During manned flight, the pilot who is positioned within the fuselage, modifies the wings orientation with respect to the fuselage by manipulating the control bar that is rigidly attached to the wing. To modify the vehicle's orientation, the pilot applies a force to the control bar to shift the wing with respect to the fuselage. By the principle of weight shift, this in turn modifies the lift profile of the wing, resulting in a new wing orientation with respect to the fuselage. 

\begin{figure}[ht]
	\centering
	\includegraphics[width=1\linewidth]{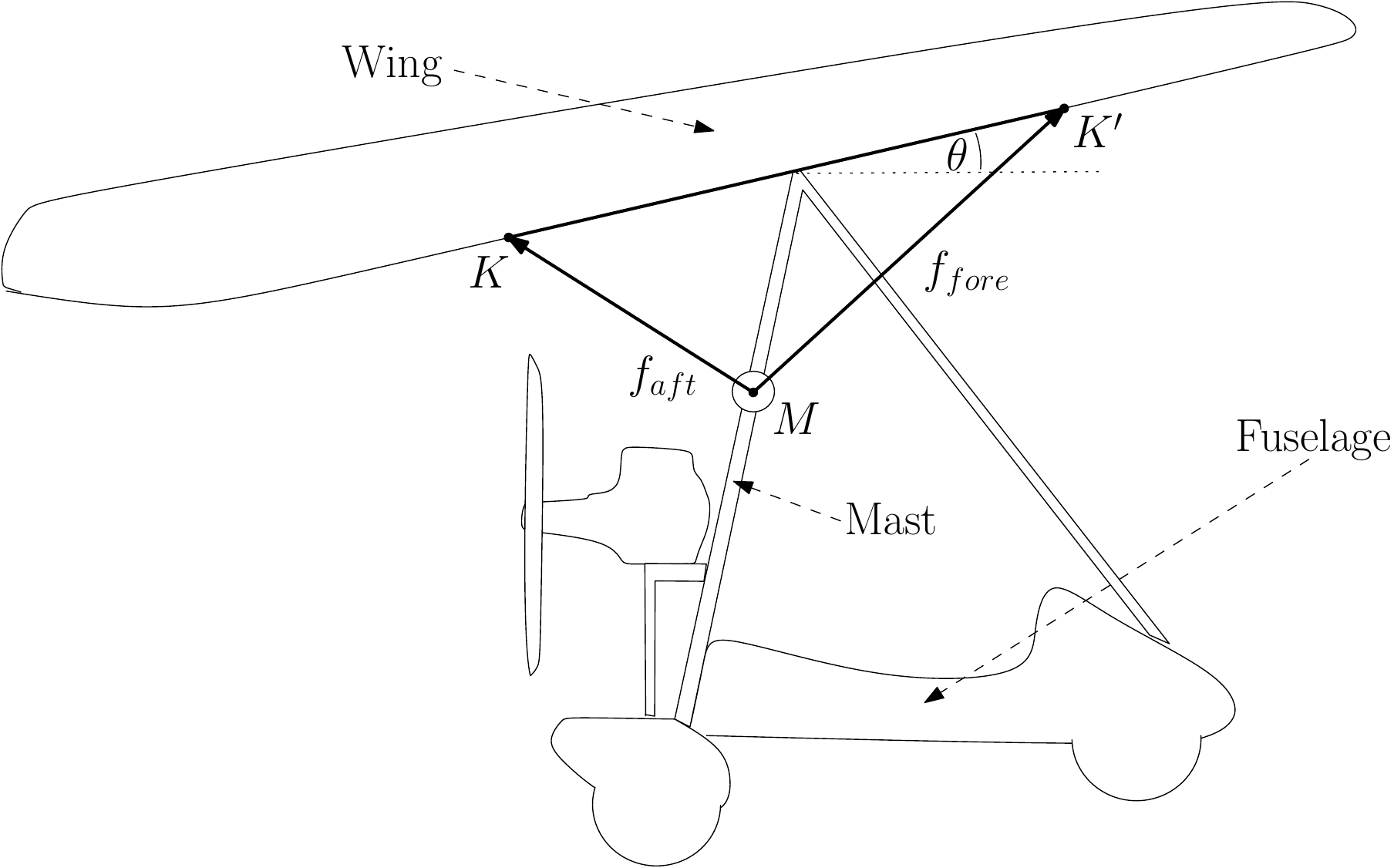}
	\caption{Schematic of weight shift kinematics of flexible-wing aircraft with two servo winches acting on wing to affect pitch.}
	\label{fig:schematic}
\end{figure}

The work presented in this paper is a first phase of a long-term project aiming at devising a model-free control algorithm for a commercial flexible-wing cargo aircraft. The first phase focuses on controlling the wing's orientation with respect to the fuselage when the aircraft is either stalled on the ground or is in a taxi mode on the runway. To this end, an experimental mock-up is set up to emulate the relative actuation mechanism between the wing and fuselage. The weight-shift kinematics of this mechanism is schematically depicted in Figure~\ref{fig:schematic}. A pair of servomotor winches are mounted at point $M$ on the mast, and connected to points $K$ and $K'$ on the keel tube. The orientation of the wing with respect to the fuselage is represented by the pair of rotation angles, $\mathbf{\theta}$ and $\mathbf{\phi}$ corresponding to pitch and roll of the wing. It is important to note that the wing does not rotate along the yaw axis, represented by $\mathbf{\psi}$, as it is constrained by the hang point on that axis. The set point of the control bar corresponds to the resting orientation of the wing in stable steady altitude and steady speed flight. Pull-only forces applied to the wing by the winch servomotors are represented by the pair of position vectors ${f_{fore}}$ and ${f_{aft}}$, and act in the free motion axis of pitch  perpendicular to the keel tube with unknown angles proportional to $\angle KK'M$ and $\angle K'KM$, respectively. For this work, it is assumed that the wing control in the pitch axis is sufficiently decoupled from the roll axis, such that the servomotor winch actions due to their geometric configuration do not induce the roll motion of the wing.  
\section{Instrumentation and Measurement Platform}
\label{sec:Inst}

\begin{figure}
\centering
\includegraphics[width=0.45\textwidth]{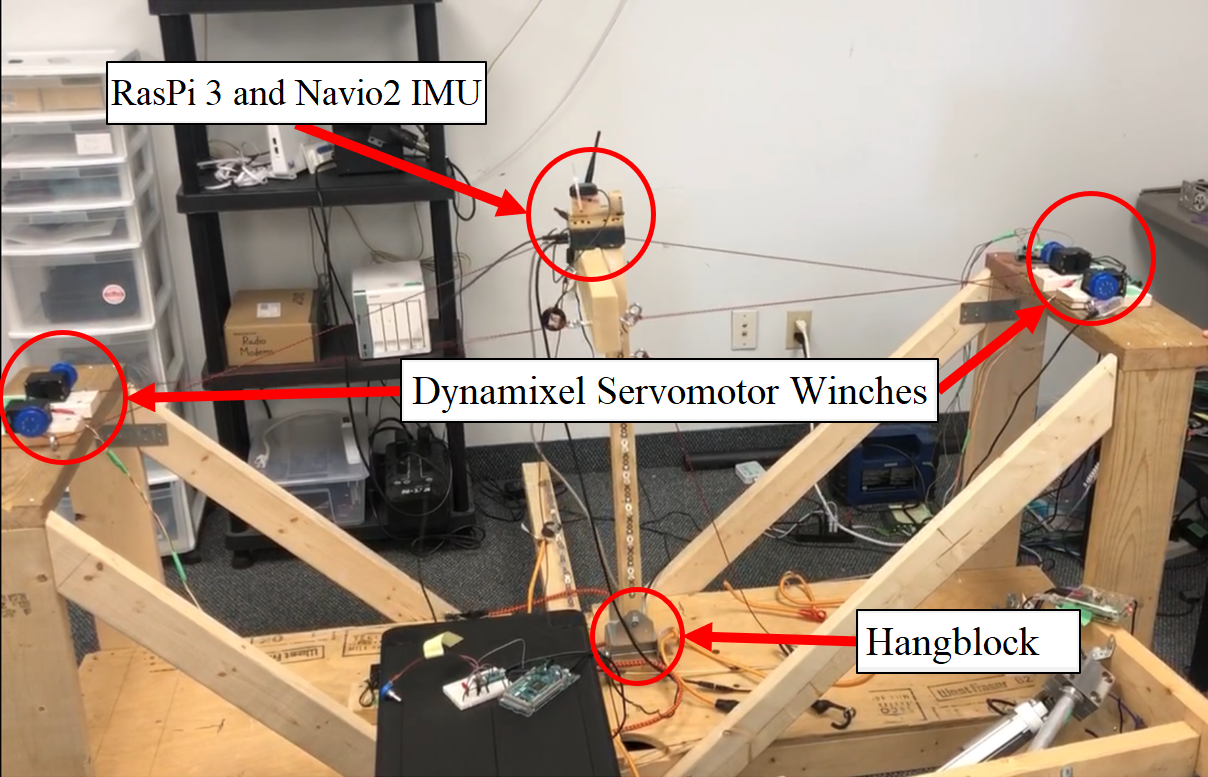}
\caption{Experimental mock-up used to emulate the motion of the fuselage about the wing in ground test. Note that the wing-mast system is oriented upside down; i.e., the top surface of the middle bar, representing the wing, is facing the floor, while the mast likewise sits above it. The figure also shows the Navio2+Raspi IMU and controller, servomotor winches, and hangblock.}
\label{fig:hardwaremockup}
\end{figure}

\begin{figure*}[ht]
	\centering
	\includegraphics[width=0.6\linewidth]{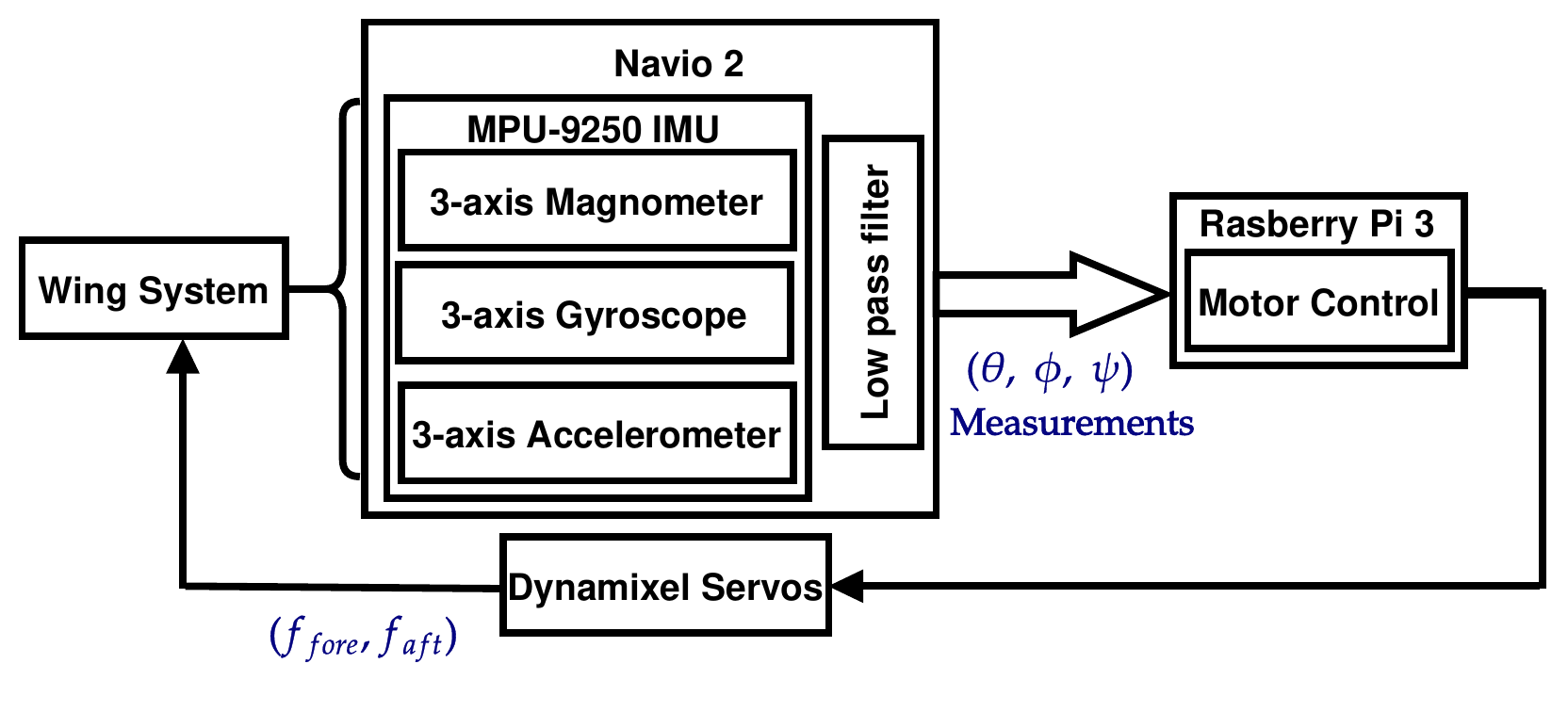}
    \caption{Flow schematic of experimental instrumentation.}
\label{fig:flowdiagram}
\end{figure*}

The experimental mock-up emulating the relative motion between the fuselage and the wing is shown in Figure~\ref{fig:hardwaremockup}. A block diagram of how the system's interconnected blocks are interfaced is revealed in Figure~\ref{fig:flowdiagram}. Efforts were made to ensure each hardware device and supporting software is either open-source or has publicly available documentation. An Emlid Navio2~\cite{emlid} board attached to a Raspberry Pi~3~\cite{raspi} is selected as the controller's computational unit due its relatively convenient and easy user setup and prevalence as a hardware of choice for open-source flight controller platforms, such as ArduPilot~\cite{Ardu} and ROS~\cite{ROS}. Navio2 provides redundant wing orientation estimates through the combination of sensor readings from either the main on-board InvenSense MPU-9250 9-Degree-of-Freedom (DOF) Inertial Measurement Unit (IMU)~\cite{TDK} or the secondary Microelectronics LSM9DS1 9 DOF IMU~\cite{STM}, as well as global positioning system (GPS). Each IMU has a three-axis magnetometer, three-axis gyroscope, and three-axis accelerometer.  The IMU signals are filtered by a programmable digital low pass filter, which is then fed to a portable mini-computer (Raspberry Pi~3 model~B). The unit has a 1.2GHz 64-bit quad-core ARMv8 CPU, 1GB of RAM and a Cortex-M3 co-processor.

Angular position measurements of the wing orientation were sampled at \SI{20}{\Hz}, a rate far below the maximum of \SI{8}{\kHz} supported by the dual IMUs. The total root mean square (RMS) noise of the MPU-9250 used for inertial measurement feedback is provided to be \SI{0.1}{\degree\per\second}. The measurements data are filtered by a low-pass filter with a cut-off frequency of \SI{250}{\Hz}. The control of the platform was actuated by two Dynamixel XM-430 servomotors~\cite{Dyna}. The servomotors receive current input commands at a rate of \SI{20}{\Hz} from the controller embedded within the Raspberry Pi~3. In reference to Figure~\ref{fig:hardwaremockup}, the system is arranged as if the aircraft wing was upside down, where the hangblock and wing are closest to the ground, and the mast sits above it. This setup was chosen for its convenience but it should have no effect on the validity of the experimental results. The mock-up allows for the mast to move relative to the wing. The servomotor winches are rigidly attached to the wing keel, and the attachment points are at opposite points along the mast. The results presented in Section~\ref{sec:simulation} are captured from experiments with this mock-up system where two servomotors are mounted directly to the wing. Desired reference position commands are programmed prior to the experiments. The learning algorithm decides the best control policy using IMU observations, in real time, and transmits current signals, based torque commands, to the servomotors to track the reference signal.

\begin{figure*}[ht]
	\centering
	\includegraphics[width=0.7\linewidth]{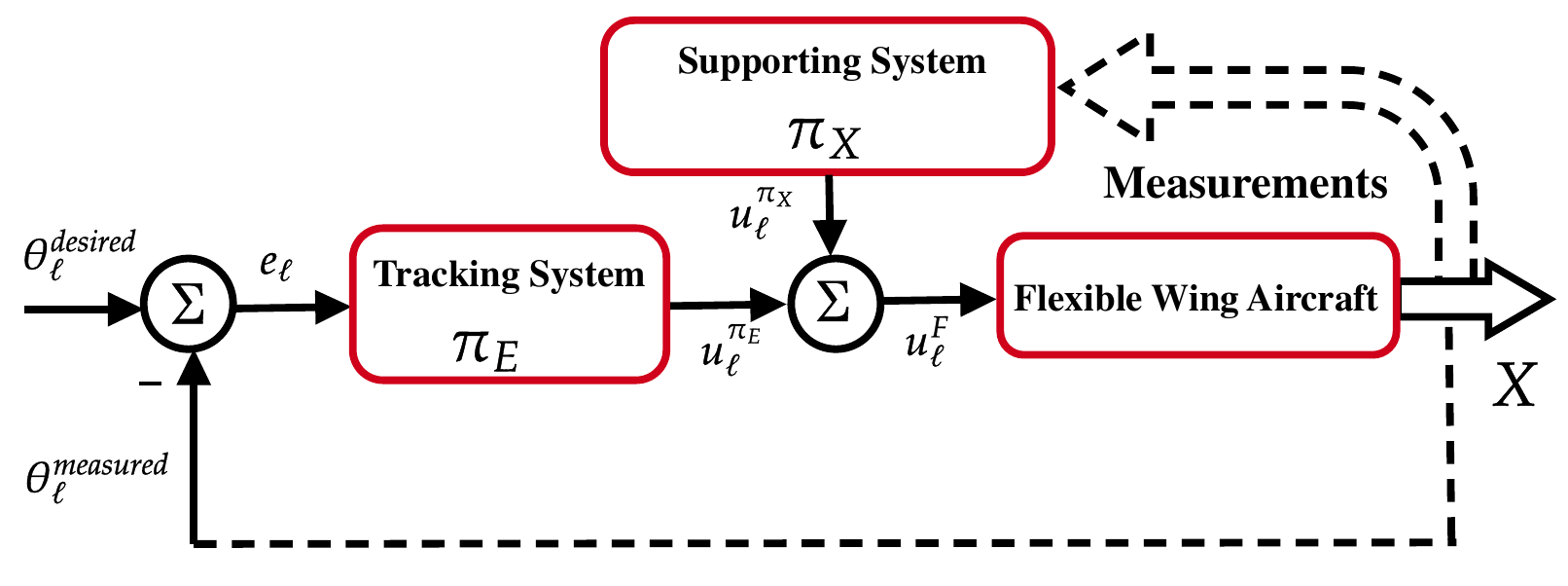}
	\caption{Feedback control loop.}
	\label{fig:controller}
\end{figure*}

\section{The Control and Optimization Problem }
\label{sec:control}
In the following section, ideas behind the online machine learning process are first developed. Thereafter, the governing Bellman or temporal difference equations are derived. Further, this section discusses how the different control tasks (i.e., tracking and stabilization) are coordinated simultaneously in real-time.

\subsection{Optimal Control Structure}
Due to the aircraft's complex nonlinear dynamics, it is necessary to avoid building a control or computational approach that relies explicitly on existence of an aerodynamic model. As such, any control solution would better be based on robust model-free computational mechanisms. 

The wing's pitch control mechanism is realized using two interacting control objectives.
The first looks for a control strategy that receives real-time tracking error signals, arranged using pre-designed criteria, and provides a control signal that minimizes the tracking error. The second searches for an auxiliary control policy that supports the stability of the overall system during the maneuvers.
%
%
The full autonomous control system is schematically represented in Figure~\ref{fig:controller}. These control objectives are integrated together and implemented using a guided search algorithm that is developed herein based on an innovative machine learning process (as shall be discussed later).

The structure of the tracking error vector $\displaystyle E_{\ell} \in \mathbb{R}^o$ reflects the design objectives corresponding to the tracking control strategy $\pi_E$ which in turn decides the tracking control signal $u^{\pi_E}_\ell \in \mathbb{R}^m$. The error vector $\displaystyle E_{\ell}$ relies on various dynamic forms of the tracking error signals $e_\ell$ ($\ell$ is a time-index). On another side, the stabilizing control policy $\pi_X$ maps the observable key measurements in vector $X_\ell \in \mathbb{R}^n$ to a state feedback control signal $u_\ell^{\pi_X} \in \mathbb{R}^s$ in order to support the system's stability during the navigation process. The overall control $u^F_\ell \in \mathbb{R}^{\max\{s,m\} }$ which is applied to the actuation system, results from combining the dynamical effects from both control policies  (i.e., $\pi_E$ and $\pi_X$). The control signals $u^{\pi_E},u^{\pi_X},$ and hence $u^{F}$, are the actuation signals responsible to move the control bar of the flexible-wing aircraft. It is worth to note that in order to generalize the proposed approach, the control signals $u^{\pi_E}$ and $u^{\pi_X}$ could have different vector sizes and the total control signal $u^F$ sums the dynamical effects of the matched actuation signals in both control laws. In this work, the control signals which are fed to the actuation systems are scalars. This simplifies the form of the collective control signal $u^F$.

Performance indices $J^{\pi_E}(E_\ell)$ and $J^{\pi_X}(X_\ell)$ are used to asses the usefulness of the attempted policies ${\pi_E}$ and ${\pi_X}$, respectively, so that 
\begin{eqnarray}
\nonumber
J^{\pi_E}(\boldsymbol{E}_\ell)&=&\displaystyle \sum_{i=\ell}^{\infty} C^E\left(\boldsymbol{E}_i,{\boldsymbol{u}}_i^{\pi_E}\right) \\ 
J^{\pi_X}(\boldsymbol{X}_\ell)&=&\displaystyle \sum_{i=\ell}^{\infty} C^X\left(\boldsymbol{X}_i,{\boldsymbol{u}}_i^{\pi_X}\right),
\end{eqnarray} 
where $C^E$ and $C^X$ are cost functions associated with the performance indices $J^{\pi_E}(\boldsymbol{E}_\ell)$ and $J^{\pi_X}(\boldsymbol{X}_\ell)$, respectively. They are given by   
\begin{eqnarray}
\nonumber
C^E(\boldsymbol{E}_\ell,\boldsymbol{u}_\ell^{\pi_E})=\frac{1}{2}\left(\boldsymbol{E}^T_\ell \, \boldsymbol{Q}^E \, \boldsymbol{E}_\ell+{\boldsymbol{u}}_\ell^{\pi_E \, T} \, {\boldsymbol{R}}^E \, {\boldsymbol{u}}_\ell^{\pi_E}\right) \\
C^X(\boldsymbol{X}_\ell,\boldsymbol{u}_\ell^{\pi_X})=\frac{1}{2}\left(\boldsymbol{X}^T_\ell \, \boldsymbol{Q}^X \, \boldsymbol{X}_\ell+{\boldsymbol{u}}_\ell^{\pi_X \, T} \, {\boldsymbol{R}}^X \, {\boldsymbol{u}}_\ell^{\pi_X}\right), 
\nonumber
\end{eqnarray}
where $\boldsymbol{Q}^E \in \mathbb{R}^{o \times o}, \, \boldsymbol{R}^E \in \mathbb{R}^{m \times m}, \boldsymbol{Q}^X \in \mathbb{R}^{n \times n},$ and $\boldsymbol{R}^X \in \mathbb{R}^{s \times s} >0$ are symmetric positive definite weighting matrices.

The objective functions $C^E$ and $C^X$ have convex forms and quadratic dependencies on the different policies and real-time measurements. These forms motivate and enable Lyapunov stability proofs for the underlying temporal difference solutions~\cite{AbouheafCDC13,AbouheafCH2014}.

\textbf{Remark 1:} The design of the optimization problem could variate from the one that is considered herein. This is decided by the designer which judges the different segments of the optimization problem and how they are hierarchically organized. In this development, it depends on the way key measurements are allowed and collected for the control design and on the dynamics which the controller needs to consider or regulate during the navigation process.

\subsection{Mathematical Solution Framework}
The control solution developed herein maps the optimization objectives mentioned above into temporal difference forms using a discrete-time optimal control framework~\cite{Lewis_2012}. Quadratic forms of solving value functions $V^{\pi_E}(\boldsymbol{E}_\ell)$ and $V^{\pi_X}(\boldsymbol{X}_\ell)$ are advised to evaluate the quality of the computed control strategies. They are motivated based on the structures of the indices $J^{\pi_E}(\boldsymbol{E}_\ell)$ and $J^{\pi_X}(\boldsymbol{X}_\ell)$ and the associated cost functions $C^E$ and $C^X$, respectively, such that   
\begin{eqnarray*}
V^{\pi_E}(\boldsymbol{E}_\ell)=\displaystyle \frac{1}{2} \boldsymbol{E}_\ell^T \, \boldsymbol{S}^E \, \boldsymbol{E}_\ell &\equiv&  J^{\pi_E}(\boldsymbol{E}_\ell)\\
V^{\pi_X}(\boldsymbol{X}_\ell)=\displaystyle \frac{1}{2} \boldsymbol{X}_\ell^T \, \boldsymbol{S}^X \, \boldsymbol{X}_\ell  &\equiv& J^{\pi_X}(\boldsymbol{X}_\ell),
\end{eqnarray*}
where $S^E \in \mathbb{R}^{o\times o}$ and $S^X\in \mathbb{R}^{n \times n}$ are solution matrices. They play a major role in the guided search policies found using the online adaptive learning process.  

The matrices $S^E$ and $S^X$ are computed using a temporal difference solution framework in real-time, and hence are relevant to the choice of the best policies ${\pi_E}$ and ${\pi_X}$, respectively, using the interactive learning process.

The value functions $V^{\pi_E}$ and $V^{\pi_X}$ are utilized to form Bellman equations (temporal difference solution structures) for the underlying control mechanism, such that
\begin{eqnarray}
\nonumber
V^{\pi_E}(\boldsymbol{E}_\ell)=\displaystyle C^E(\boldsymbol{E}_\ell,\boldsymbol{u}_\ell^{\pi_E}) + V^{\pi_E}(\boldsymbol{E}_{\ell+1})\\
V^{\pi_X}(\boldsymbol{X}_\ell)=\displaystyle C^X(\boldsymbol{X}_\ell,\boldsymbol{u}_\ell^{\pi_X}) +V^{\pi_X}(\boldsymbol{X}_{\ell+1}).
\label{Bell}
\end{eqnarray}
These Bellman equations indicate that two interacting optimization processes are solved simultaneously, where it is required to drive the tracking error $\boldsymbol{E}$ to zero and, at the same time, optimize the dynamics $\boldsymbol{X}$ along the navigation trajectories. Further, Bellman equations \eqref{Bell} enable the integration between two environments. The first is related to solving the navigation control problem, while the second enables an approximate dynamic programming solution (i.e., machine learning solution) for the problem. The online adaptive learning control process starts with an initial control strategy for each control decision and then the learning process, employing the above Bellman equations, directs the control strategies (i.e., learn better control decisions) using a value iteration process which is guaranteed to converge.

Herein, the error vector $\boldsymbol{E}_\ell$ is structured as follows: 
\[\boldsymbol{E}_\ell=
\begin{bmatrix}
 e_\ell & e_{\ell-1} & e_{v\ell} & e_{v\ell-1} & e_{s\ell} & s_{s\ell-1}
\end{bmatrix}^T
\]
where $e_{v\ell}$ and $e_{v\ell-1}$  are error derivatives, with respect to time, evaluated at time $\ell$ and $\ell-1$, respectively. $e_{s\ell}$ and $s_{s\ell-1}$ are the moving averages calculated as $e_{s\ell}=\displaystyle \frac{1}{N}\sum_{i=\ell-N}^{\ell} e_i$ and $e_{s\ell-1}=\displaystyle \frac{1}{N}\sum_{i=\ell-N-1}^{\ell-1} e_i$, respectively.

The way the error vector $\boldsymbol{E}_\ell$ is calculated, combines many sub-objectives which are optimized together. It minimizes the local tracking error $e_\ell$ while considering its previous instance $e_{\ell-1}$. Further, it smoothens the control decision by looking backward in-time to include error derivatives evaluated at time instances $\ell$ and $\ell-1$, as well as evaluating average of the errors across longer time-intervals $N$. It is worth to mention that the vector formulation $\boldsymbol{E}_\ell$ enables more advanced forms of equivalent discrete Proportional-Integral-Derivative (PID) structures (i.e., dependence on $e_\ell, e_{\ell-1},$ and $e_{\ell-2}$) or even considers higher-order derivatives and integral equivalents (i.e., dependence on $e_\ell, e_{\ell-1},e_{\ell-2},\dots,$ and $e_{\ell-N}$).

The solution of the optimal tracking problem relies on solving a number of differential equations, a subgroup of which is solved offline and at the same time they do not allow dynamical forms of the error signals. Herein, the formulation of the adaptive learning mechanism allows for an online solution as well as using a variety of tracking error dynamical forms.

The experimental setup, described in Section~\ref{sec:Inst}, provides measurements related to the aerodynamic orientation of the wing. This made it convenient to choose the states $\boldsymbol{X}_\ell$ as
\[\boldsymbol{X}_\ell=
\begin{bmatrix}
\theta_\ell & \theta_{v\ell} & \theta_{a\ell} 
\end{bmatrix}^T,
\]
where $\theta_\ell, \theta_{v\ell},$ and $\theta_{a\ell}$ are the pitch attitude, pitch velocity, and pitch acceleration, respectively.

The desired control policies ${\boldsymbol{\pi}_E}$ and ${\boldsymbol{\pi}_X}$ are linear feedback polices which are used to decide on the different control signals, such that
\begin{eqnarray}
\boldsymbol{u}^{\pi_E}_\ell= {\boldsymbol{\pi}_E} \, \boldsymbol{E}_\ell,\quad  \boldsymbol{u}^{\pi_X}_\ell= {\boldsymbol{\pi}_X} \, \boldsymbol{X}_\ell,
\end{eqnarray} \newline
Next, we will explain how to compute the real-time negative feedback control strategies $\boldsymbol{\pi}_E$ and $\boldsymbol{\pi}_X$.

\section{Machine Learning Platform}
\label{sec:Machine}

The online computational machine learning framework is driven by the design of the control problem. First, we will present a computational platform based on an innovative value iteration implementation for an online reinforcement learning solution. Then, adaptive critics are employed to provide neural network implementation for the approximate dynamic programming solution.

\subsection{Guided Search Process}
The policies $\boldsymbol{\pi}_E$ and $\boldsymbol{\pi}_X$ are guided toward the intended optimization objectives stated by the designer as follows: 
\begin{eqnarray}
\boldsymbol{u}^{\pi_E}= - \boldsymbol{P}^E \frac{\partial V^{\pi_E}(\boldsymbol{E}_\ell)}{\partial \boldsymbol{E}_\ell}, \quad
\boldsymbol{u}^{\pi_X}= - \boldsymbol{P}^X \frac{\partial V^{\pi_X}(\boldsymbol{X}_\ell)}{\partial \boldsymbol{X}_\ell},
\label{eq:consig}
\end{eqnarray}
where $\boldsymbol{P}^E \in \mathbb{R}^{m\times o}$ and $\boldsymbol{P}^X\in \mathbb{R}^{s\times n}$ are guiding search vectors which carry the intentions and dynamic preferences of the control mechanism designer.

\noindent The vectors $\boldsymbol{P}^E$ and $\boldsymbol{P}^X$ contain selective dynamic forms (i.e., introduced by the designer to reflect their objectives regarding the entries in vectors $\boldsymbol{E}$ and $\boldsymbol{X}$) as will be highlighted in the experimental analysis. This in turn promotes flexibility for a new class of online learning processes with guided search features.

The control polices $\boldsymbol{\pi}_E=- \boldsymbol{P}^E \, \boldsymbol{S}^E$ and $\boldsymbol{\pi}_X=- \boldsymbol{P}^X \, \boldsymbol{S}^X$  (i.e., these associated with $\displaystyle \frac{\partial V^{\pi_E}(\boldsymbol{E}_\ell)}{\partial \boldsymbol{E}_\ell}$ and $\displaystyle \frac{\partial V^{\pi_X}(\boldsymbol{X}_\ell)}{\partial \boldsymbol{X}_\ell}$) are applied in real-time using a value iteration process as will be explained next.

\subsection{Value Iteration Algorithm}
An online reinforcement learning solution is developed using the aforementioned Bellman equations~\eqref{Bell}. The solution is realized using a two-step value iteration process. The first updates the solving value functions (i.e., $\boldsymbol{S}^E$ and $\boldsymbol{S}^X$) using (\ref{Bell}) while the second extracts the new or improved policies (i.e., $\boldsymbol{\pi}_E$ and $\boldsymbol{\pi}_X$). The online value iteration process is detailed out in Algorithm~\ref{alg:ValueIteration}. Bellman equations provide the temporal difference platform necessary to guide the online learning process beyond the initially selected polices $\boldsymbol{\pi}_E^0$ and $\boldsymbol{\pi}_X^0$. Note that, unlike policy iteration paradigms,  these initial policies do not need to be admissible. Hence, the intentionally guided policies are improved along the trajectory of the aircraft. 

\textbf{Remark 2:} Value iteration processes are proven to converge in general for single and multi-agent systems based on Lyapunov stability approaches if the underlying systems are stabilizable~\cite{Abouheaf2014,AbouheafCH2014,Land1996}. This is mainly due to the properties of convex objective cost functions (i.e., $C^E$ and $C^X$ in this work) and consequently Bellman solution forms (i.e.,~\eqref{Bell}). Hence, the value functions evolve in a bounded manner, such that
  \begin{eqnarray}
    &0\le V^{\pi_E(0)}(\boldsymbol{E}_\ell) \le V^{\pi_E(1)}(\boldsymbol{E}_\ell)\le\dots \le \nonumber \\ &V^{\pi_E(t)}(\boldsymbol{E}_\ell)\le \dots \le V^{\pi_E(*)}(\boldsymbol{E}_\ell), \nonumber \\
    &0\le V^{\pi_X(0)}(\boldsymbol{X}_\ell) \le V^{\pi_X(1)}(\boldsymbol{X}_\ell)\le  \dots \le \nonumber \\ &V^{\pi_X(t)}(\boldsymbol{X}_\ell)\le \dots \le V^{\pi_X(*)}(\boldsymbol{X}_\ell),
  \label{seq}
  \end{eqnarray}
  where the value functions $V^{\pi_E(*)}$ and $V^{\pi_X(*)}$ are the optimal response solutions for Bellman equations \eqref{Bell}. In other words, the tracking error and motion dynamics are stable if the aircraft system is controllable.

The online value iteration learning process guides the solving value functions toward the best values $V^{\pi_E(*)}(\boldsymbol{E}_\ell)$ and $V^{\pi_X(*)}(\boldsymbol{X}_\ell)$, and hence best guided policies ${\boldsymbol{\pi}_E^*}$ and ${\boldsymbol{\pi}_X^*}$.

\subsection{Adaptive Critics Implementation}
\label{sec:critics}
Adaptive critics are employed as neural network approximation tools for the introduced online reinforcement learning solution~\cite{Widrow1973}. The adaptive critics scheme is implemented using means of neural network structures, namely actor-critic networks, for each optimization control process. They provide guided search solutions for the underlying Bellman equations. The critic and actor structures approximate the solving value functions and the associated policies in an interactive manner implicitly during the iterative update of underlying Bellman equations. The actor network reflects the improvements in the guided control strategy while its quality is approximated by the critic network. The weights of the actor-critic structures are adapted using a gradient decedent approach motivated by the linear forms of the control policies.

The value functions $V^{\pi_E}$ and $V^{\pi_X}$ (i.e., the critic structures) are approximated so that 
\begin{eqnarray}
\nonumber
\hat V^{\pi_E}(\boldsymbol{E}_\ell)=\displaystyle \frac{1}{2} \boldsymbol{E}_\ell^T\, \boldsymbol{\Omega}_c^E \,{\boldsymbol{E}_\ell},
\hat V^{\pi_X}(\boldsymbol{X}_\ell)=\displaystyle \frac{1}{2} \boldsymbol{X}_\ell^T\, \boldsymbol{\Omega}_c^X \,{\boldsymbol{X}_\ell},
\label{crit}
\end{eqnarray}
where $\boldsymbol{\Omega}_c^E \in \mathbb{R}^{o \times o}$ and $\boldsymbol{\Omega}_c^X \in \mathbb{R}^{n \times n}$ are the critic approximation weights of the value functions $\hat V^{\pi_E}(\boldsymbol{E}_\ell)$ and $\hat V^{\pi_X}(\boldsymbol{X}_\ell)$, respectively. The critic network forms are motivated by the structures of the value functions $V^{\pi_E}$ and $V^{\pi_X}$ respectively. 

In a similar fashion, the guided search polices (i.e., the actor structures) are approximated so that 
\begin{equation}
 \boldsymbol{\hat u}^{\pi_E}_\ell= \boldsymbol{\Omega}_a^{E} \boldsymbol{E_\ell},  \quad \boldsymbol{\hat u}^{\pi_X}_\ell= \boldsymbol{\Omega}_a^{X} \boldsymbol{X_\ell},
\label{actr}
\nonumber
\end{equation}
where $\boldsymbol{\Omega}_a^{E} \in \mathbb{R}^{m \times o}$ and $\boldsymbol{\Omega}_a^{X} \in \mathbb{R}^{s \times n}$ are the actor approximation weights of the policies ${\boldsymbol{\pi}_E}$ and ${\boldsymbol{\pi}_X}$, respectively.

The different approximation weights are updated using a gradient descent approach that applies a minimization criteria on the squared approximation errors. 
The approximation errors of the critic networks have squared forms, so that
\begin{eqnarray}
\nonumber
\varepsilon_c^{E_\ell}=\frac{1}{2}\left(\hat V^{\pi_E}(\boldsymbol{E}_\ell)-\hat V^{T_E}(\boldsymbol{E}_\ell)\right)^2,
\\
\varepsilon_c^{X_\ell}=\frac{1}{2}\left(\hat V^{\pi_X}(\boldsymbol{X}_\ell)-\hat V^{T_X}(\boldsymbol{X}_\ell)\right)^2,
\label{criterr}
\end{eqnarray}
where the target values $\hat V^{T_E}(\boldsymbol{E}_\ell)$ and $\hat V^{T_X}(\boldsymbol{X}_\ell)$ are calculated using 
\begin{eqnarray}
\nonumber
\hat V^{T_E}(\boldsymbol{E}_\ell)=\displaystyle C^E(\boldsymbol{E}_\ell,\boldsymbol{\hat u}_\ell^{\pi_E}) + \hat V^{\pi_E}(\boldsymbol{E}_{\ell+1})\\ \nonumber
\hat V^{T_X}(\boldsymbol{X}_\ell)=\displaystyle C^X(\boldsymbol{X}_\ell,\boldsymbol{\hat u}_\ell^{\pi_X}) +\hat V^{\pi_X}(\boldsymbol{X}_{\ell+1}).
\label{crittarget}
\end{eqnarray}
The critic weights are adapted according to a gradient descent rule such that 
\begin{eqnarray}
	\nonumber
	\boldsymbol{\Omega}_c^{E(t+1)}=\boldsymbol{\Omega}_ c^{E(t)}  - \alpha_c^E \left[\frac{\partial\varepsilon_c^{E_\ell}}{\partial {\boldsymbol{\Omega}_ c^{E}}}\right]^{(t)}
	\\ 
		\boldsymbol{\Omega}_c^{X(t+1)}=\boldsymbol{\Omega}_ c^{X(t)}  - \alpha_c^X \left[\frac{\partial\varepsilon_c^{X_\ell}}{\partial {\boldsymbol{\Omega}_ c^{X}}}\right]^{(t)},
	\label{crittune}
\end{eqnarray}
where $0<\alpha_c^E, \alpha_c^X<1$ are critic networks learning rates, $\displaystyle \frac{\partial\varepsilon_c^{E_\ell}}{\partial {\boldsymbol{\Omega}_ c^{E}}}= \left(\hat V^{\pi_E}(\boldsymbol{E}_\ell)-\hat V^{T_E}(\boldsymbol{E}_\ell)\right) \boldsymbol{E}_\ell \, \boldsymbol{E}_\ell^T,$
$\displaystyle  \frac{\partial\varepsilon_c^{X_\ell}}{\partial {\boldsymbol{\Omega}_ c^{X}}}= \left(\hat V^{\pi_X}(\boldsymbol{X}_\ell)-\hat V^{T_X}(\boldsymbol{X}_\ell)\right) \boldsymbol{X}_\ell \, \boldsymbol{X}_\ell^T,$ and $t$ refers to the iteration update index.

Similarly, the squared approximation errors for the actor networks are defined by
\begin{eqnarray}
\nonumber
\varepsilon_a^{E_\ell}=\frac{1}{2}\left( \boldsymbol{\hat u}^{\pi_E}_\ell- \boldsymbol{\hat u}^{T_E}_\ell\right)^2,\quad
\nonumber
\varepsilon_a^{X_\ell}=\frac{1}{2}\left( \boldsymbol{\hat u}^{\pi_X}_\ell- \boldsymbol{\hat u}^{T_X}_\ell\right)^2,
\label{acterr}
\end{eqnarray}
where the target values $\boldsymbol{\hat u}^{T_E}_\ell$ and $\boldsymbol{\hat u}^{T_X}_\ell$ are calculated as follows
$$
{\boldsymbol{\hat u}^{T_E}_\ell}= - \boldsymbol{P}^E \, \boldsymbol{\Omega}_c^{E} \boldsymbol{E}_\ell, \quad
{\boldsymbol{\hat u}^{T_X}_\ell}= - \boldsymbol{P}^X \, \boldsymbol{\Omega}_c^{X} \boldsymbol{X}_\ell.
$$
Using gradient descent, the actor network weights are adapted according to the following rule 
	\begin{eqnarray}
	\nonumber
	\boldsymbol{\Omega}_a^{E(t+1)}=\boldsymbol{\Omega}_ a^{E(t)}  - \alpha_a^E \left[\frac{\partial\varepsilon_a^E}{\partial {\boldsymbol{\Omega}_ a^{E_\ell}}}\right]^{(t)}
	\\ 
	\boldsymbol{\Omega}_a^{X(t+1)}=\boldsymbol{\Omega}_ a^{X(t)}  - \alpha_a^X \left[\frac{\partial\varepsilon_a^X}{\partial {\boldsymbol{\Omega}_ a^{X_\ell}}}\right]^{(t)},
\label{acttune}
	\end{eqnarray}
	where $0<\alpha_a^E, \alpha_a^X<1$ are actor networks learning rates, $\displaystyle \frac{\partial\varepsilon_a^{E_\ell}}{\partial {\boldsymbol{\Omega}_ a^{E}}}= \left( \boldsymbol{\hat u}^{\pi_E}_\ell-{\boldsymbol{\hat u}^{T_E}_\ell}\right)  \, \boldsymbol{E}_\ell^T,$ and 
	$\displaystyle  \frac{\partial\varepsilon_a^{X_\ell}}{\partial {\boldsymbol{\Omega}_ a^{X}}}= \left( \boldsymbol{\hat u}^{\pi_X}_\ell-{\boldsymbol{\hat u}^{T_X}_\ell}\right)  \, \boldsymbol{X}_\ell^T$.

\begin{figure}
	\centering
	\includegraphics[width=0.9\linewidth]{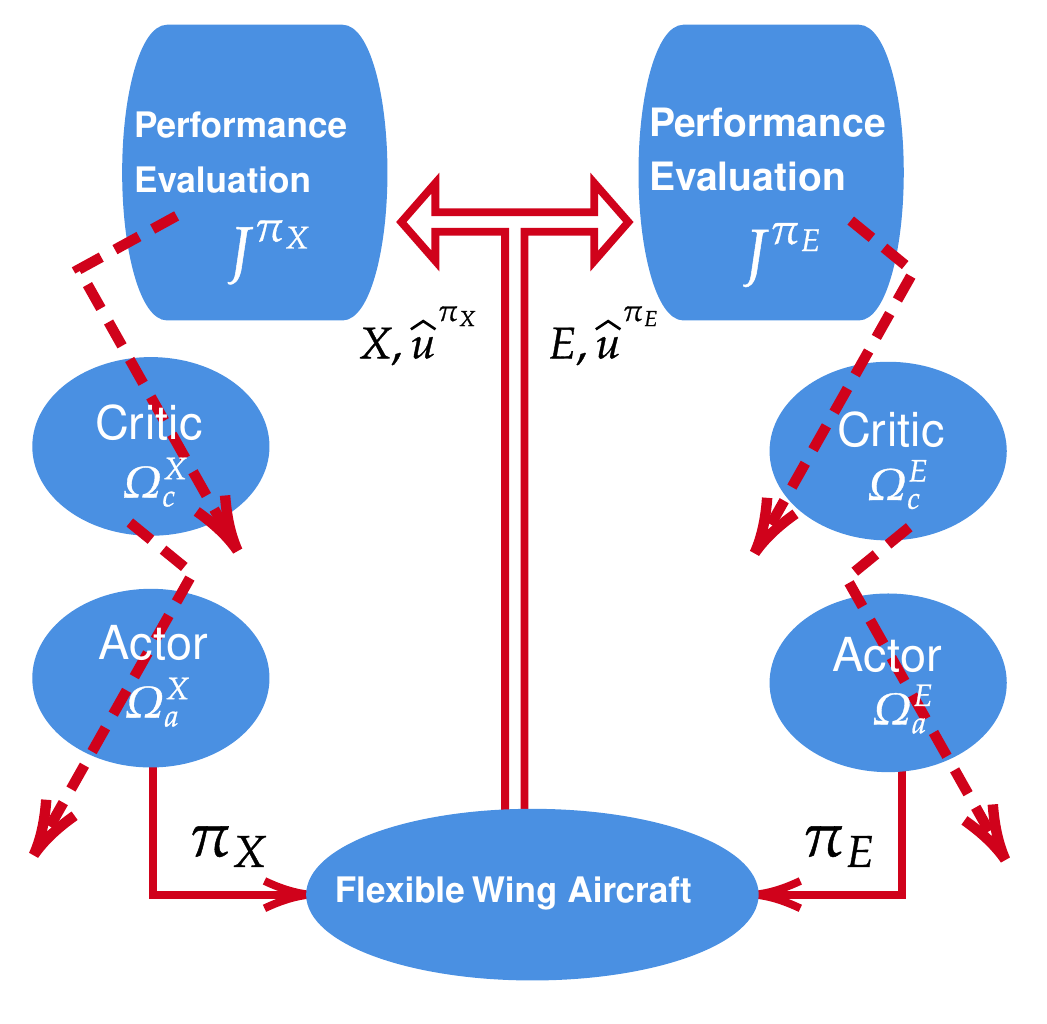}
\caption{Adaptive critics structure.}
\label{fig:actorc-ritic}
\end{figure}

A full schematic diagram of the adaptive critics process is shown in Figure~\ref{fig:actorc-ritic} for the combined control problem in hand. The different actor and critic weights are updated simultaneously in real-time following the scope of Algorithm~\ref{alg:ValueIteration}.

\begin{algorithm2e}
	\caption{\label{alg:ValueIteration} Online Value Iteration Process.}
	\DontPrintSemicolon
	\KwIn{Desired trajectory $\theta^{desired}_\ell,$ weighting matrices ${\boldsymbol{Q}}^E,{\boldsymbol{R}}^E,{\boldsymbol{Q}}^X,$ and ${\boldsymbol{R}}^X$, guiding search vectors $\boldsymbol{P}^E$ and $\boldsymbol{P}^X$, tracking error evaluation interval $N$. }
	\KwOut{Policies $\boldsymbol{\pi}_E,$ $\boldsymbol{\pi}_X,$ and tracking error ${e}_\ell,$ for $\ell=0,1,\ldots$}
	\Begin{
		$\ell=0, t = 0$ \tcc*[h]{time and strategy indices}\;
		Initialize solving matrices $\boldsymbol{S}^{E\{0\}}$ and $\boldsymbol{S}^{X\{0\}}$ \tcc*[h]{Positive definite} and hence initial policies $\boldsymbol{\pi}_E^0$ and $\boldsymbol{\pi}_X^0$ and tracking errors interval $e_{\ell-1}, \dots,e_{\ell-N-1}$\;
		Measure $\theta_\ell,\theta_{v\ell},\theta_{a\ell}$ and then calculate the errors $ e_\ell,\, e_{\ell-1},\,  e_{v\ell},\, e_{v\ell-1},\, e_{s\ell}, \, e_{s\ell-1}$\;
		\Repeat(\tcc*[h]{Training/Search loop}){Satisfactory trajectory-tracking performance (i.e., acceptable tracking error). }
		{
			Compute the different control signals $u^{\pi_E(t)}$ and $u^{\pi_X(t)}$ using (\ref{eq:consig}) \;
			Measure $\theta_{\ell+1},\theta_{v\ell+1},\theta_{a\ell+1}$ and then calculate the errors $ e_{\ell+1},\, e_{\ell},\,  e_{v\ell+1},\, e_{v\ell},\, e_{s\ell+1}, \, e_{s\ell}$\;
			Evaluate the solving value functions
			$V^{\pi_E(t+1)}(\boldsymbol{E}_\ell)=\displaystyle C^E(\boldsymbol{E}_\ell,\boldsymbol{u}_\ell^{\pi_E(t)}) + V^{\pi_E(t)}(\boldsymbol{E}_{\ell+1})$ $V^{\pi_X(t+1)}(\boldsymbol{X}_\ell)=\displaystyle C^X(\boldsymbol{X}_\ell,\boldsymbol{u}_\ell^{\pi_X(t)}) +V^{\pi_X(t)}(\boldsymbol{X}_{\ell+1}).
			$\;
			Extract the improved control policies \newline
			$
			{\boldsymbol{\pi}_E}^{t+1}= - \boldsymbol{P}^E  S^{E(t+1)}, 
			{\boldsymbol{\pi}_X}^{t+1}= - \boldsymbol{P}^X  S^{X(t+1)},
			$\;
			$\ell\leftarrow \ell+1$ \tcc*[h]{Update real-time index}\;
			$t\leftarrow t+1$\tcc*[h]{Update policy index}\;

		}
	}
\end{algorithm2e}

\subsection{Complexity of the Learning Mechanism}
Algorithm~\ref{alg:ValueIteration} provides multiple merits concerning the complexity and scalability of the navigation or tracking problems. First, the learning process can be conditioned for any number of tracking signals simultaneously (i.e., for a large tracking or navigation process that contains many tracking objectives, that would be easy to implement compared to solving a high number of differential equations in an offline mode). Second, the tracking error signals could involve many dynamical forms (i.e., moving velocity, acceleration, moving average, etc) which were not possible or easy-to-implement before. Third, this scheme allows for coupled optimized dynamical processes (i.e., possible optimization of multiple coupled and interactive dynamical objectives for multiple problems without the need to independently solve each problem).  Finally, the arrangement of Bellman equations enables simple and straightforward adaptive critics solution, since the control strategies hold linear forms.

\section{Experimental Results}
\label{sec:simulation}
The online model-free adaptive learning mechanism is validated in a real-time environment using a two-servomotor pitch actuation system. The system generates servomotor pull actions dependent on the current-measured pitch position and the desired trajectory steering the wing. For the sake of experiments presented below, the trajectory reference is chosen to be sinusoidal (i.e., continuous pitch up/down commands). 

\subsection{Learning Environment}
The learning parameters of the system are selected in order to reflect the physical constraints of the measured dynamical parameters and also encode the control design preferences that prioritize the desired system response. The weighting matrices capture the physical limitations of the dynamical parameters (i.e., the states $\theta_\ell, \theta_{v\ell}, \theta_{a\ell},e_\ell$ and the actuation control signal $u^F$). They are chosen as $R^E= 10^{-7},$   $R^X= 10^,$\\
$Q^E= 10^{-4} \begin{bmatrix}
25 & 0 & 0 & 0 & 0 & 0\\
0 & 25 & 0 & 0 & 0 & 0\\
0 & 0 & 0.25 & 0 & 0 & 0\\
0 & 0 & 0 & 0.25 & 0 & 0\\
0 & 0 & 0 & 0 & 25 & 0\\
0 & 0 & 0 & 0 & 0 & 25
\end{bmatrix}$, \\ $Q^X= 10^{-6} \begin{bmatrix}
25& 0  & 0\\
0 & 25 & 0\\
0 & 0 & 0.0025
\end{bmatrix}.$

\noindent Vectors $P^E$ and $P^X$ for the guided search policies $\pi^E$ and $\pi^X$ are selected so that
$P^E= \begin{bmatrix}
200 & 50 & 10 & 5 & 10 & 5
\end{bmatrix}$ and $P^X= \begin{bmatrix}
10 & 10 & 5
\end{bmatrix}$. The vector $P^E$  assigns more weight to minimizing the recently measured tracking errors $e_\ell$ and $e_{\ell-1}$, as opposed to the error velocity terms $e_{v\ell}$ and $e_{v\ell-1}$, to smooth down the nonlinearity transitions. Furthermore, the influences of the moving average tracking errors  $e_{s\ell}$ and $e_{s\ell-1}$ on the overall performance are weighted similarly to those of the error velocities. The vector $P^X$ reflects the gradual dynamic importance of the pitch attitude $\theta_\ell$ and the pitch velocity $\theta_{v\ell}$ over the angular acceleration $\theta_{a\ell}$. The actor and critic learning rates are selected to be small enough in order to match the actor and critic adjustments in a smooth manner. They are set to $\alpha^E_c=0.01,$ $\alpha^X_c=0.01,$ $\alpha^E_a=0.01,$ and $\alpha^X_a=0.05$. The desired navigation trajectory is an independent sinusoidal reference  signal that takes the form $\theta^{desired}_\ell= 20    \,  \sin(0.132\, \pi \, \ell)\deg$.

\subsection{Test Scenarios}
\begin{figure}[!ht]
\centering
\subfloat[]{\includegraphics[width=0.45\textwidth]{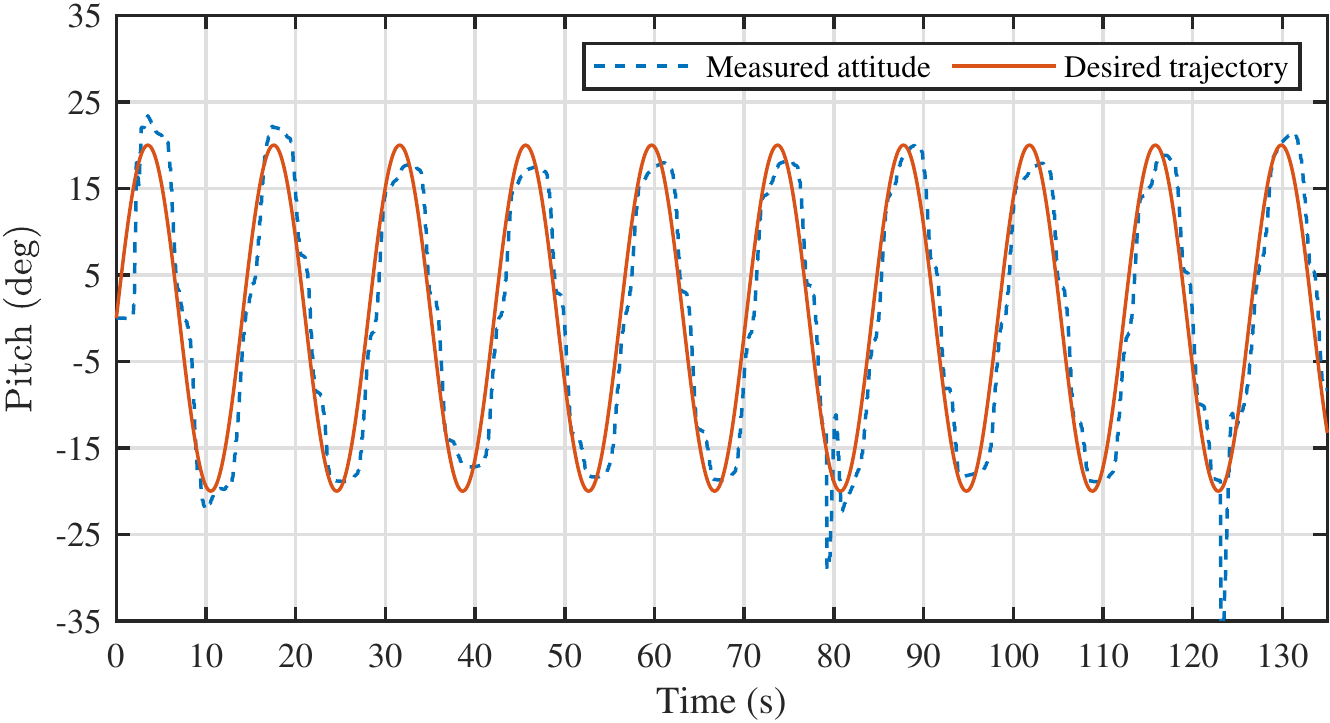}%
\label{fig:AttSim1a}}
\hfil
\subfloat[]{\includegraphics[width=0.45\textwidth]{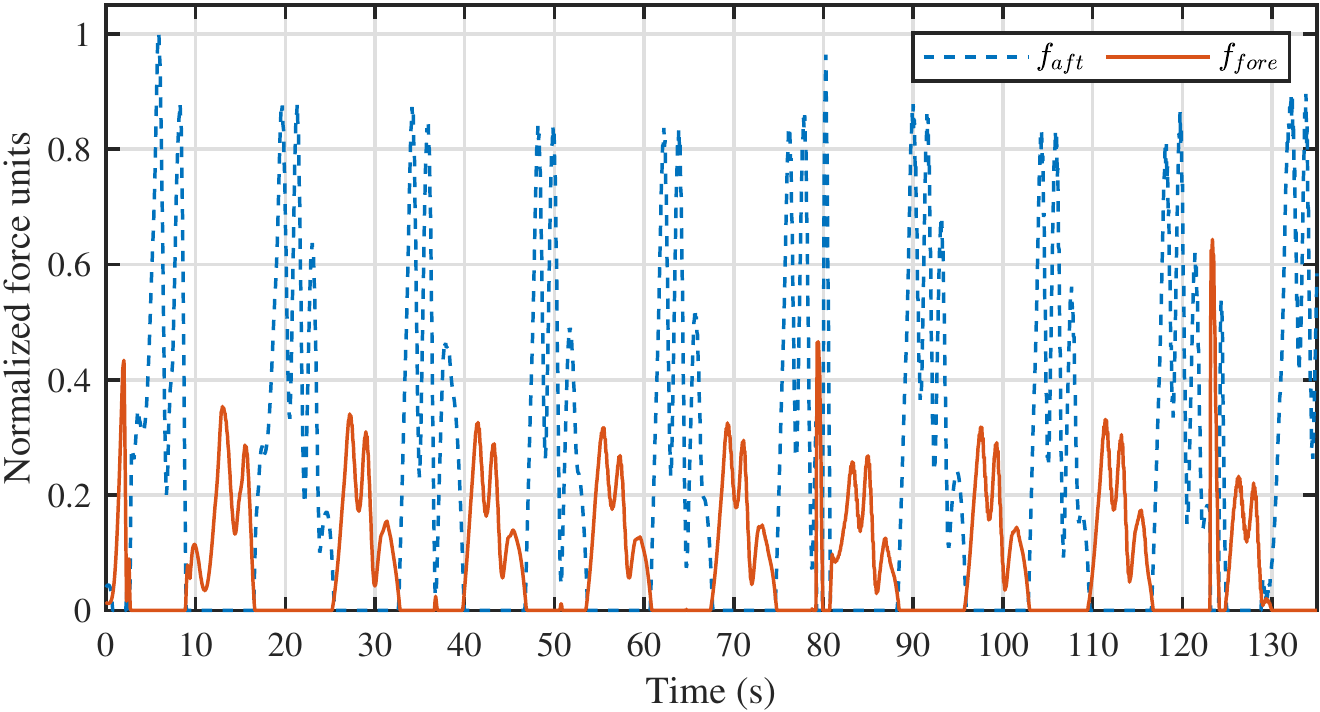}%
\label{fig:AttSim1b}}
\caption{Control performance during online learning process: (a) measured vs. desired pitch attitude, (b) acting forces on the wing's keel, $f_{fore}$ and $f_{aft}$.}
\label{fig:AttSim1}
\end{figure}

\begin{figure*}[!ht]
\centering
\subfloat[]{\includegraphics[width=0.45\textwidth]{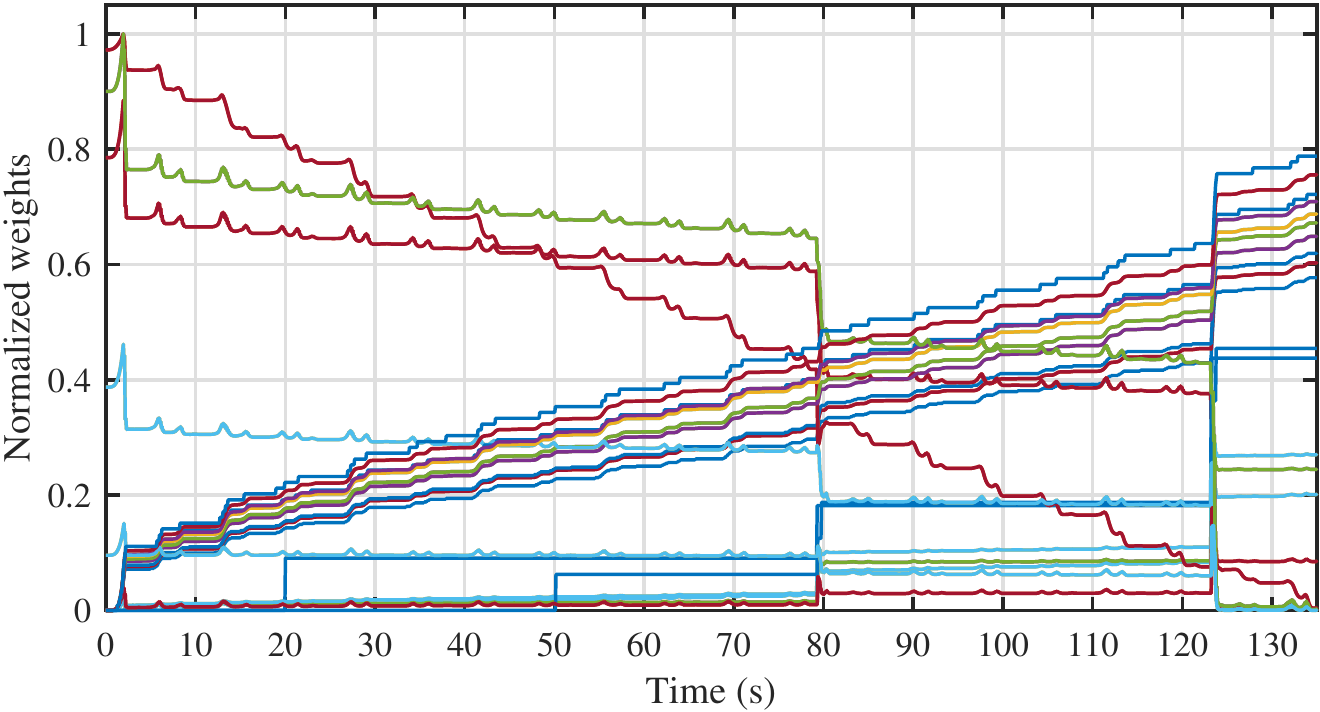}%
\label{fig:AttSimCT}}
\hfil
\subfloat[]{\includegraphics[width=0.45\textwidth]{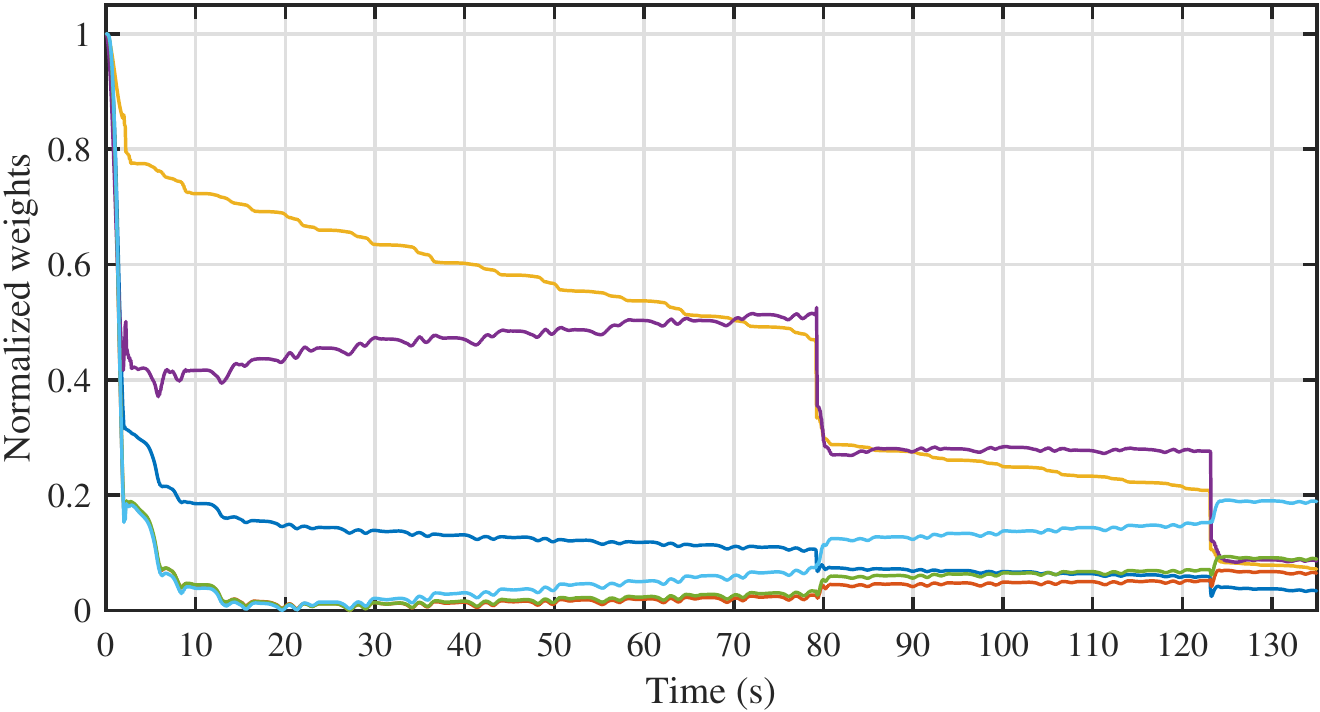} %
\label{fig:AttSimAT}}
\hfil
\subfloat[]{\includegraphics[width=0.45\textwidth]{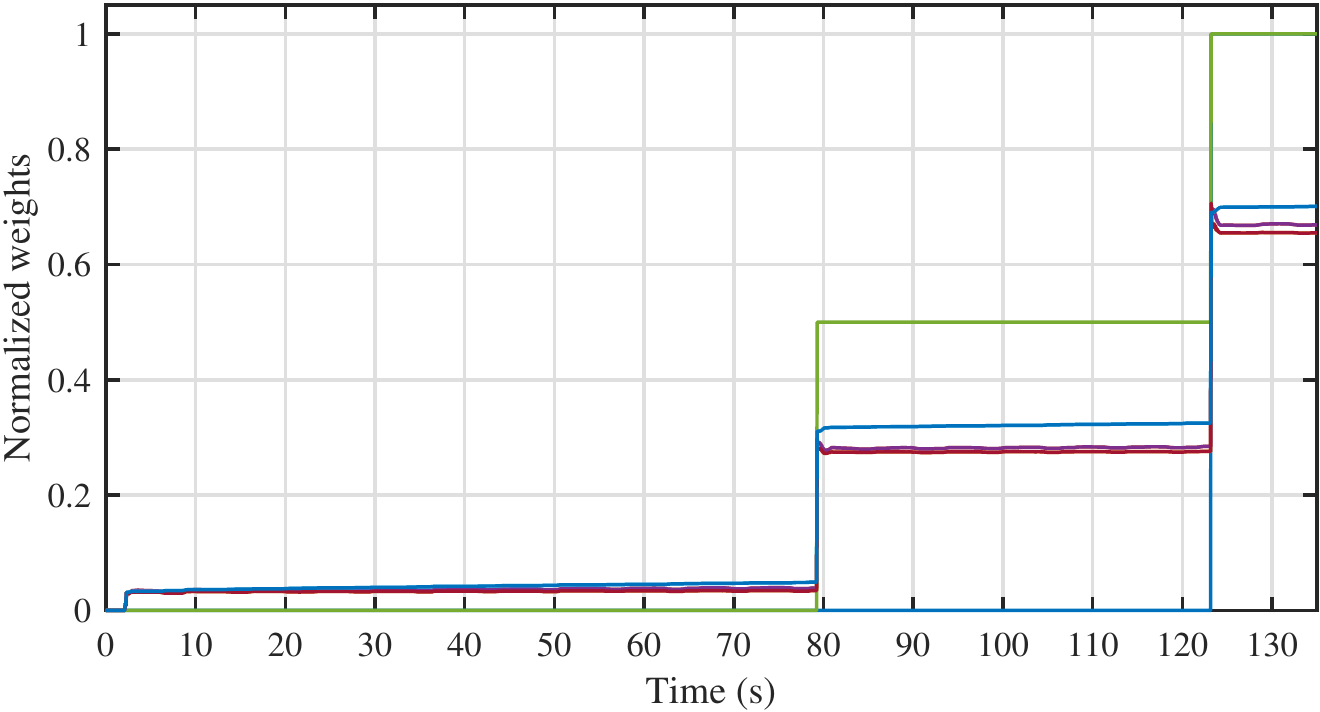}%
\label{fig:AttSimCS}}
\hfil
\subfloat[]{\includegraphics[width=0.45\textwidth]{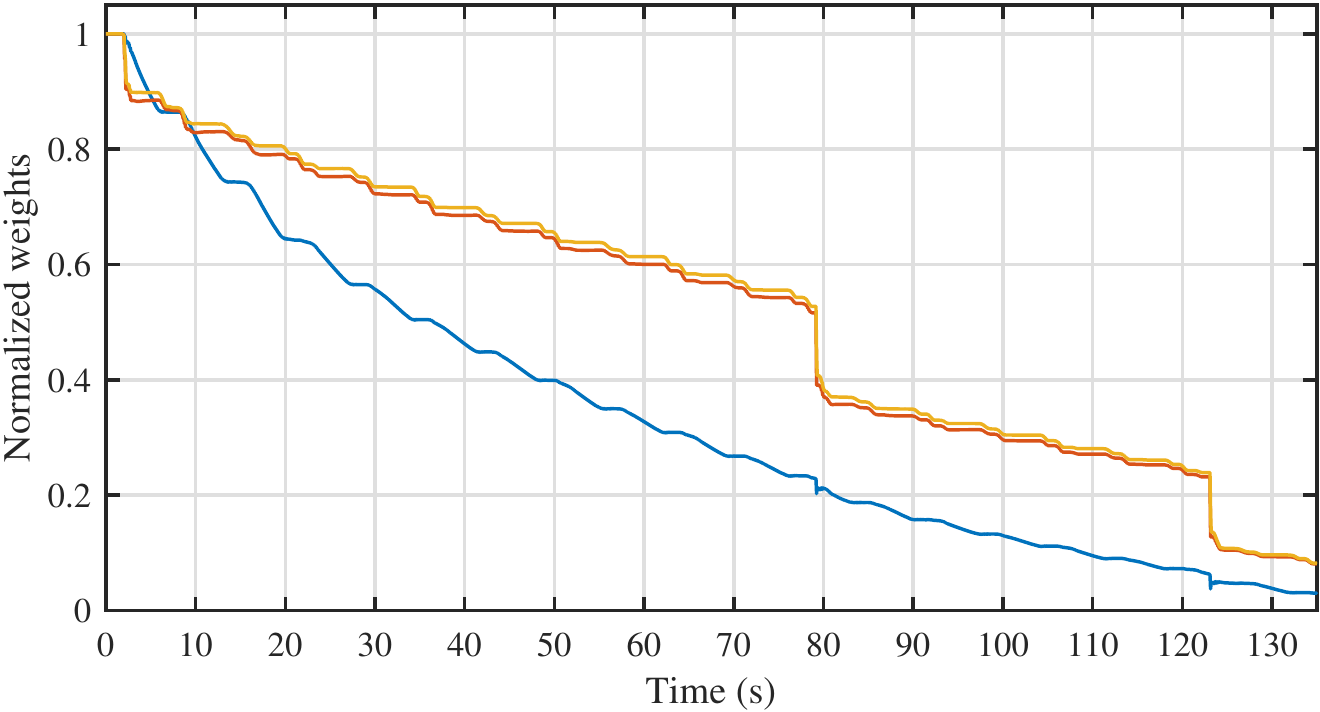}%
\label{fig:AttSimAS}}
\hfil
\caption{Evolution of actor-critic weights during learning process: (a) tracking critic unit, (b) tracking actor unit, (c) stabilizing critic unit, (d) stabilizing actor unit. }
\label{fig:ACSim1}
\end{figure*}

\begin{figure}
\centering
\includegraphics[width=0.45\textwidth]{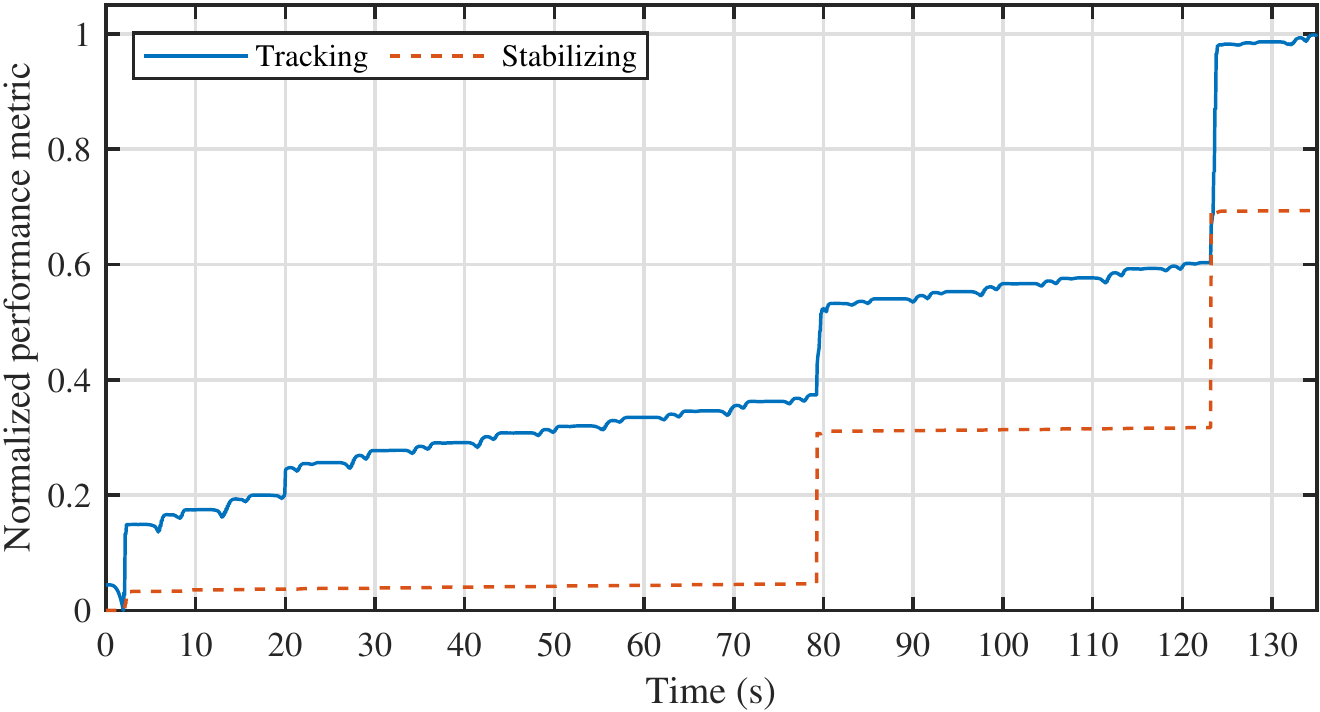}
\caption{Evolution of the policy evaluation metric.}
\label{fig:performance}
\end{figure}

The adaptive learning controller is validated using three test scenarios. 
In the first scenario, the system is tested under nominal circumstances where no external disturbance is applied on the system. 
The dynamic performance of the proposed adaptive learning controller is shown in Figure~\ref{fig:AttSim1}. This experiment reveals that, the learning algorithm successfully converges to a stable control policy which prescribes correct servomotor pull-forces for the observed system state (i.e., pitch attitude). The absolute average tracking error over the test period is found to be ${0.45} \, \deg$. It is worth noticing that the convergence behavior of the proposed control mechanism is not affected by the occasional sudden erroneous readings in the feedback measurements, as revealed at around time instants \SI{80}{\s} and \SI{123}{\s} in Figure~\ref{fig:AttSim1a}. Such measurement errors are due to some imperfections in the sensing units which could not be overcame by the adopted low-pass filter. The convergence of the adaptive learning approach is clear in Figure~\ref{fig:ACSim1}, where the actor and critic weights converge throughout the online learning process. The overall normalized forces generated by the servomotors are shown in Figure~\ref{fig:AttSim1b}. Finally, the convergence of the value iteration approach, as described in Section~\ref{sec:Machine}, is highlighted in Figure~\ref{fig:performance}. It is shown that, despite the sensor reading spikes, a general bounded non-decreasing convergence pattern is observed, as expected for the underlying value iteration process.

\begin{figure}[!ht]
\centering
\subfloat[]{\includegraphics[width=0.45\textwidth]{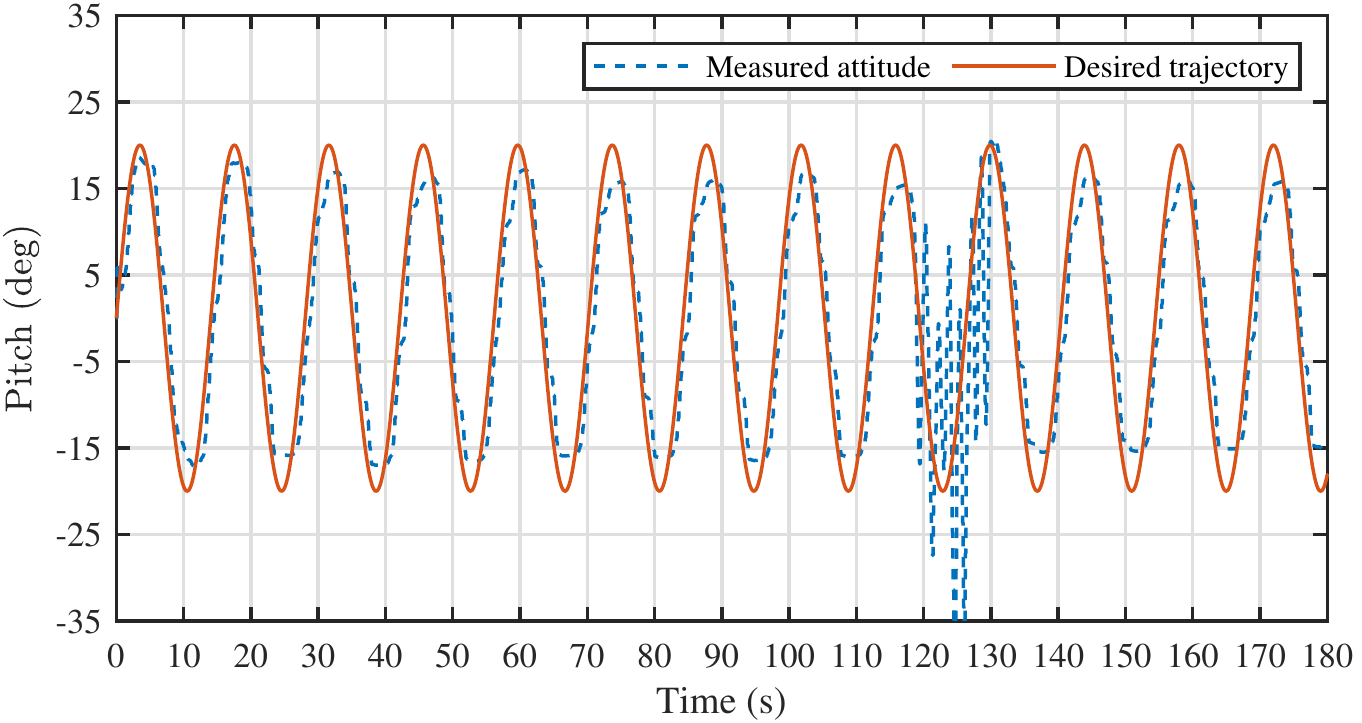}%
\label{fig:AttSim2a}}
\hfil
\subfloat[]{\includegraphics[width=0.45\textwidth]{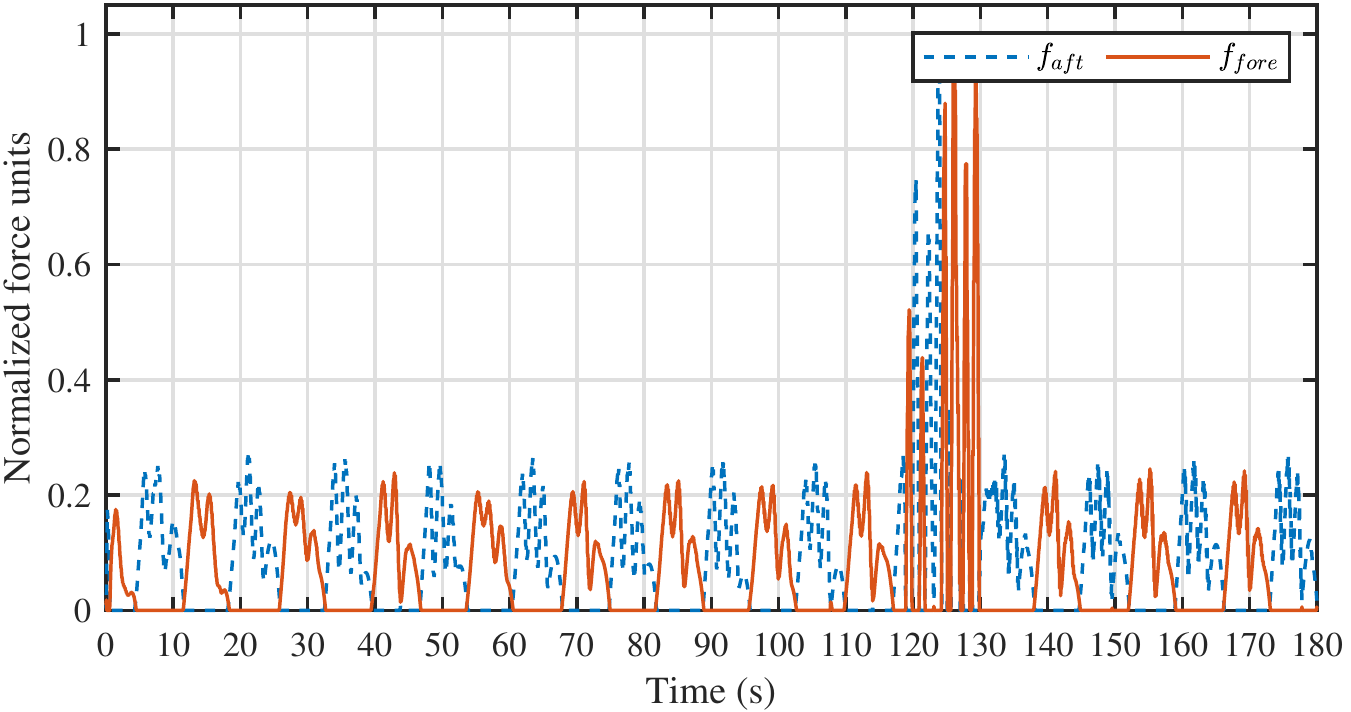}%
\label{fig:AttSim2b}}
\caption{Control performance while learning with a converged controller in the presence of mechanical disturbances: (a) measured vs. desired pitch attitude, (b) acting forces on the wing's keel, $f_{fore}$ and $f_{aft}$.}
\label{fig:AttSim2}
\end{figure}

\begin{figure}[!ht]
\centering
\subfloat[]{\includegraphics[width=0.45\textwidth]{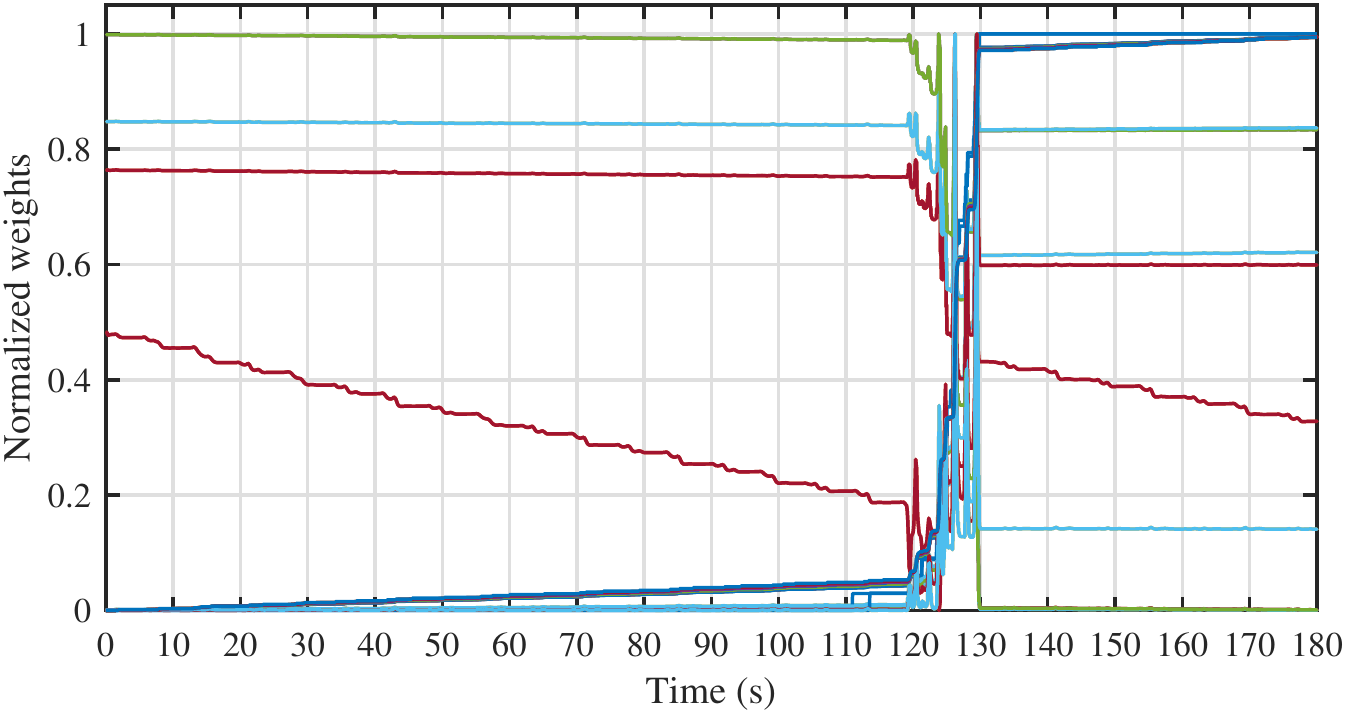}%
\label{fig:Sim2Criticstrack}}
\hfil
\subfloat[]{\includegraphics[width=0.45\textwidth]{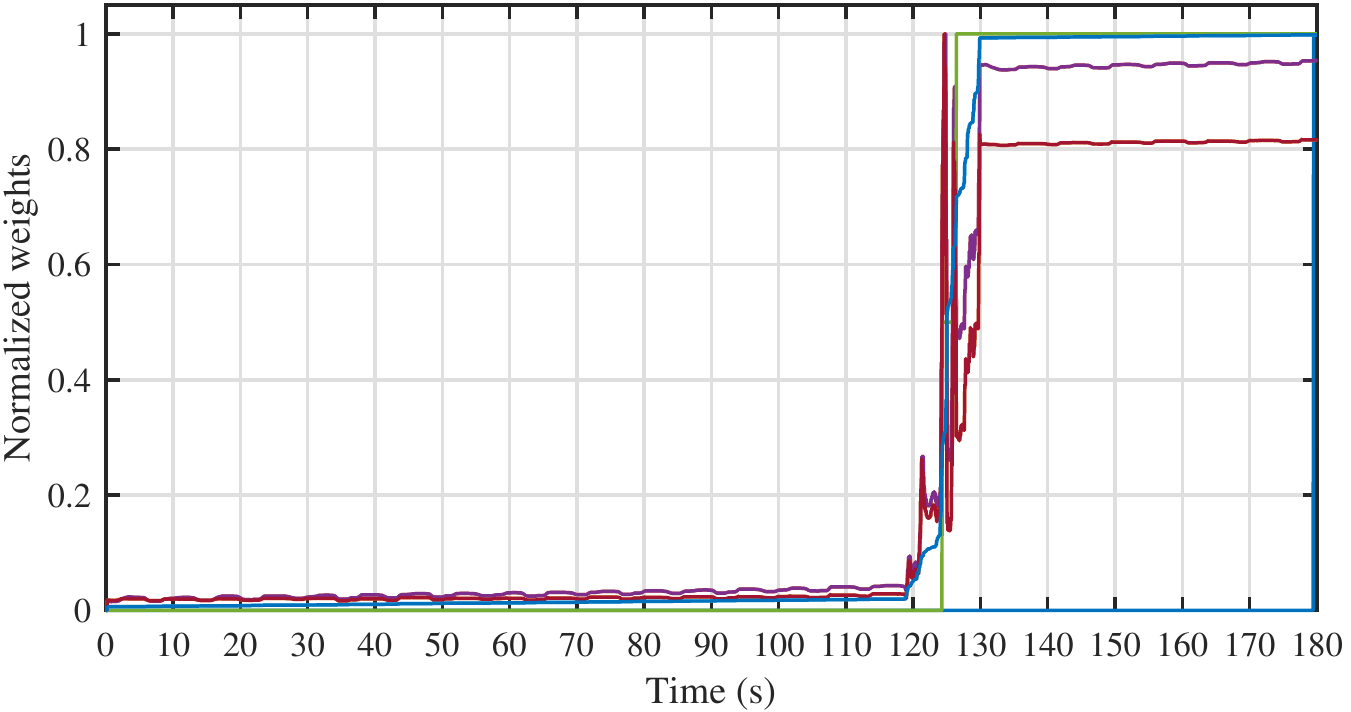}%
\label{fig:Sim2Criticsstab}}
\caption{Variations in critic weights in the face of mechanical disturbances: (a) tracking unit, (b) stabilizing unit.}
\label{fig:Sim2Critics}
\end{figure}

The second scenario is set up to demonstrate the ability of the controller to reject sudden external and forced disturbances.
The test case is a repetition of the first, except that this time, a sudden disturbance was applied at around \SI{120}{\s} for about \SI{10}{\s} by violently jerking the control bar around the pitch and roll axes to stimulate the effects of wind gusts on the wing during the flight. 
The tracking performance of the proposed controller is demonstrated in Figure~\ref{fig:AttSim2a}. The controller is started with converged control gains from a previous learning episode. As such, minimal changes in the critic weights are witnessed, as shown by the critic weights evolution in Figure~\ref{fig:Sim2Critics}. During disturbance, the critic reacts appropriately to generate large abrupt motor torques to counteract the induced perturbations, as shown in Figure~\ref{fig:AttSim2b}.
Once the dynamics are back to their nominal conditions, after that the disturbance subsided, the controller successfully resumes the tracking of the commanded pitch.

\begin{figure}[!ht]
\centering
\subfloat[]{\includegraphics[width=0.45\textwidth]{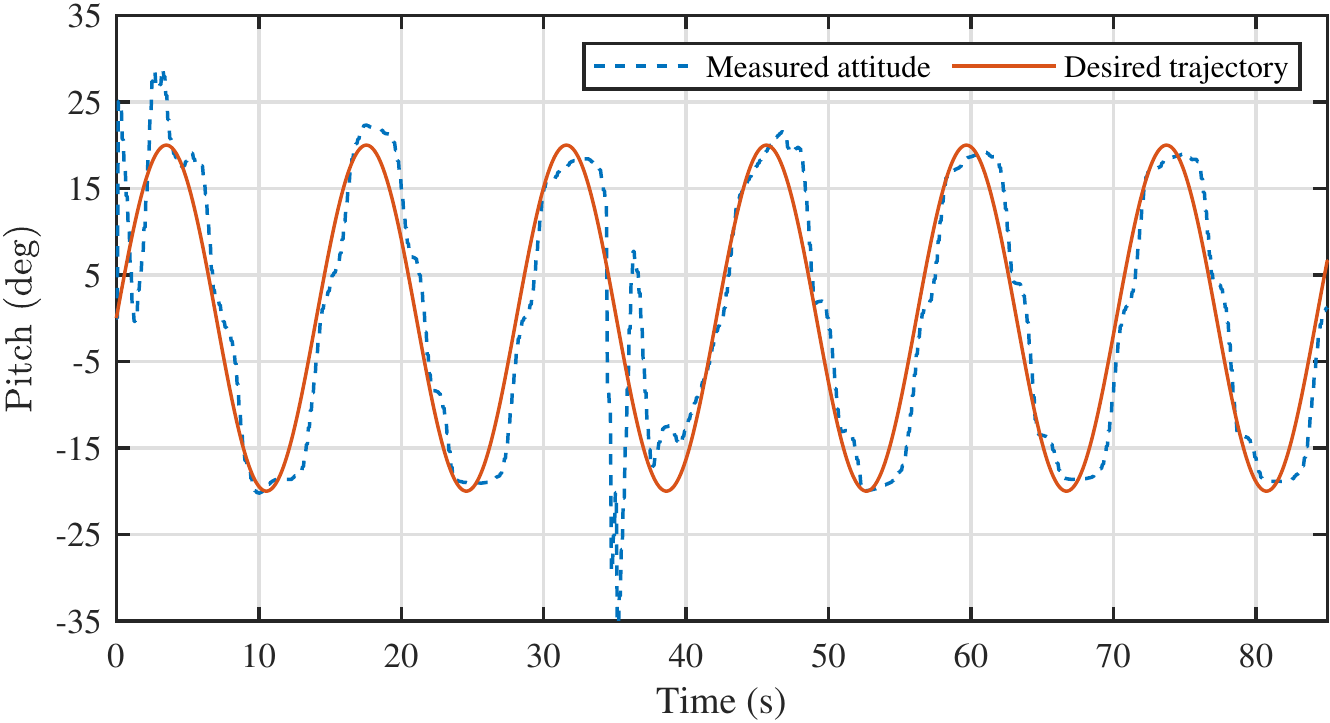}%
\label{fig:AttSim3a}}
\hfil
\subfloat[]{\includegraphics[width=0.45\textwidth]{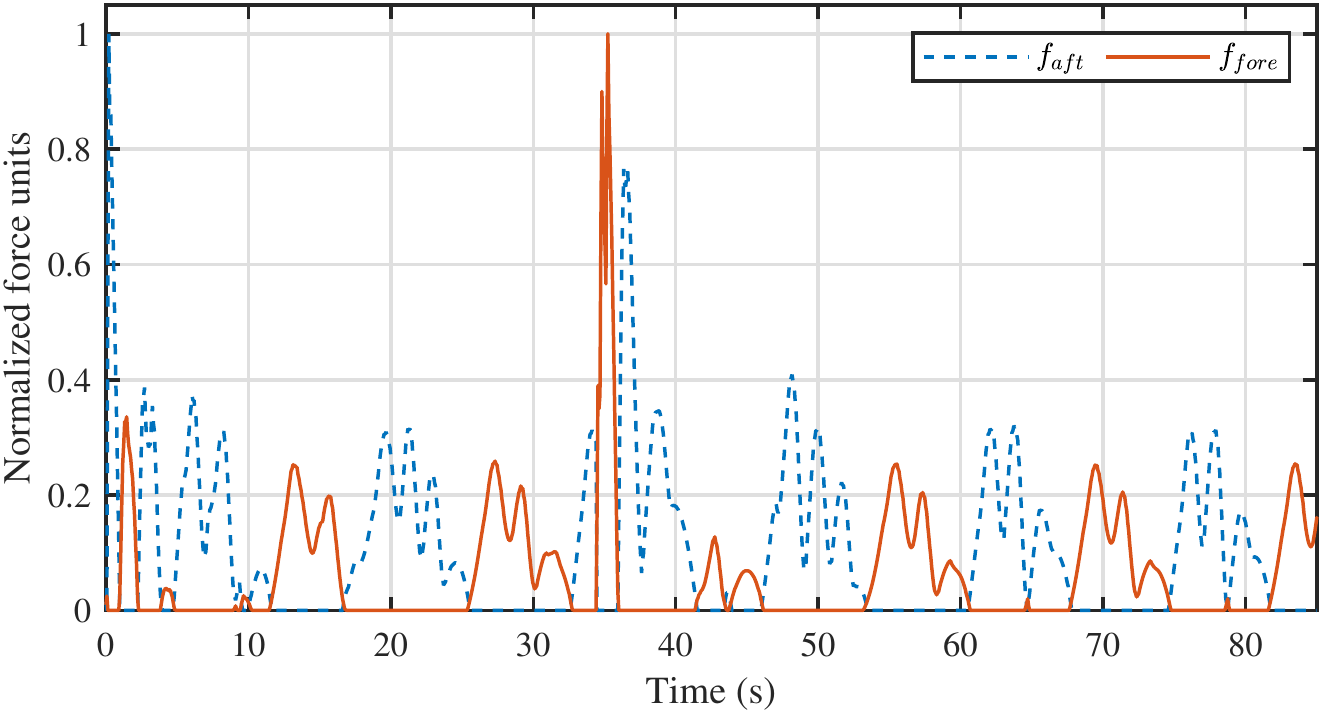}%
\label{fig:AttSim3b}}
\caption{Control performance with previously learned static control policies: (a) measured vs. desired pitch attitude, (b) acting forces on the wing's keel, $f_{fore}$ and $f_{aft}$.}
\label{fig:AttSim3}
\end{figure}

The goal of the third experiment is to show that once the controllers converge to some optimal strategies $\pi^*_E$ and $\pi^*_X$, the actor's learning mechanism can be turned off without degrading the control performance. By doing so, the controller is made to operate with static control gains. For this test case, the adopted static control policies applied are those converged from a prior experiment. The controller's performance is illustrated in Figure~\ref{fig:AttSim3}. The absolute average tracking error over the test period is found to be ${0.94} \, \deg$. This result emphasizes that the converged strategies from static ``actor-only'' controllers previously derived from the adaptive learning mechanism remain valid during the trajectory tracking process. Notice how this experiment also witnessed two abrupt spikes in the sensor measurements.

Unlike classical Q-leaning processes which employ multiple offline training episodes before settling on suitable control strategies, herein the proposed approach showed successful outcomes following a single real-time learning episode as shown in Figures~\ref{fig:AttSim1a},~\ref{fig:AttSim2a}. Additionally, the learning process exhibits capability to handle large mechanical systems. This explains the superior features of the proposed approach to follow desired trajectory reaching bounded error when some converged polices are utilized to run the system as shown in Figure~\ref{fig:AttSim3a}. The maximum observed absolute average tracking error was confined to under ${1}\, \deg$ for such mechanical actuation system (without overlooking the sensory error biasing).

When experimenting with real systems (as opposed to simulation), it is typical to notice differences between the reference and the actual signals. The differences may be due to several reasons, including noise, delay caused by digital filters, and backlashes. In this particular work, the differences also come from the coupling between the roll and pitch motions. The two were assumed to be decoupled in the first-order linear approximation of the system. However, in a real system, such as the one we adopted, they are never completely decoupled. Nevertheless, it is clear from the figures that the lag between the reference and the actual signals is bounded within an acceptable margin.


\textbf{Remark 3:} As explained earlier, the stability of the value iteration solution was investigated and proved, provided that the system under consideration is stabilizable and convex objective cost function is adopted. Figures~\ref{fig:AttSim1a},~\ref{fig:AttSim2a}, and~\ref{fig:AttSim3a} emphasize the bounded stability features and the ability of the online learning mechanism to retrieve stability under significant mechanical disturbances, noise, and false readings. These figures show that the tracking errors are bounded and non-increasing. This is reassured by examining the actuation input signals shown in Figures~\ref{fig:AttSim1b},~\ref{fig:AttSim2b}, and~\ref{fig:AttSim3b}.

\section{Conclusion}
\label{sec:conc}

The challenging autonomous navigation of flexible-wing aircraft is solved by integrating analytical and computational solution platforms into an extensible experimental instrumentation and actuation incubator. 
The feedback loop receives the inertial measurements from standard measurement package Navio2 mounted on the wing system of an aircraft and decides, in real-time, on the best control strategies without acquiring any prior knowledge on the system's dynamical model. This enabled the integration of a powerful model-free control unit with a flexible and affordable measuring circuitry without over-complicating the sensory structures for such type of systems. The quality of the resultant systems depends mostly on the design of the model-free learning process rather than the precision of the sensors. This could be faced if a different model-based strategy is followed or even complicated augmented control structures are considered for this type of aircraft. This in turn, opens the door to generalize this approach for problems with similar interests. On another side, an online guided search mechanism based on a value iteration process is introduced where the learning process decides on the real-time control strategies based on the dynamic selection of the goals of the optimization problem. The experimental results coincided with the stated objectives, where the wing is subjected to severe disturbances without destabilizing the system.

\bibliographystyle{IEEEtran}
\bibliography{Bib/mybibliography}

\end{document}